\documentclass[12pt,preprint]{aastex}

\shorttitle{Binary Stars in Cas OB6}
\shortauthors{Hillwig et al.}

\begin{document}

\received{}
\accepted{}

\title{Binary and Multiple O-Type Stars in the Cas OB6 Association}

\author{Todd C. Hillwig\altaffilmark{1,2}, Douglas R. Gies\altaffilmark{1}, 
William G. Bagnuolo, Jr., Wenjin Huang\altaffilmark{1}, \\
M. Virginia McSwain\altaffilmark{1,3,4}, and David W. Wingert\altaffilmark{1}}

\affil{Center for High Angular Resolution Astronomy and \\
Department of Physics and Astronomy, Georgia State University, 
P. O. Box 4106, Atlanta, GA 30302-4106;
todd.hillwig@valpo.edu, gies@chara.gsu.edu, 
bagnuolo@chara.gsu.edu, huang@chara.gsu.edu,
mcswain@astro.yale.edu, wingert@chara.gsu.edu}

\altaffiltext{1}{Visiting Astronomer, Kitt Peak National Observatory,
National Optical Astronomical Observatory, which is operated by the
Association of Universities for Research in Astronomy, Inc. (AURA)
under cooperative agreement with the National Science Foundation.}
\altaffiltext{2}{Current address: Department of Physics and Astronomy, 
Valparaiso University, Valparaiso, IN 46383}
\altaffiltext{3}{Current address: Department of Astronomy, 
Yale University, P. O. Box 208101, New Haven, CT 06520-8101}
\altaffiltext{4}{NSF Astronomy and Astrophysics Postdoctoral Fellow}

\slugcomment{}
\paperid{63881}


\begin{abstract}

We present the results of time-resolved spectroscopy of 13 O-type stars
in the Cas OB6 stellar association.  We conducted a survey for
radial velocity variability in search of binary systems, which
are expected to be plentiful in young OB associations.
Here we report the discovery of two new single-lined binaries, 
and we present new orbital elements for three double-lined
binaries (including one in the multiple star system HD~17505).  
One of the double-lined systems is the eclipsing binary 
system DN~Cas, and we present a preliminary light curve analysis
that yields the system inclination, masses, and radii. 
We compare the spectra of the single stars and the 
individual components of the binary stars with model synthetic 
spectra to estimate the stellar effective temperatures, 
gravities, and projected rotational velocities. 
We also make fits of the spectral energy distributions to 
derive $E(B-V)$, $R=A_V/E(B-V)$, and angular diameter. 
A distance of 1.9~kpc yields radii that are consistent with 
evolutionary models.  We find that 7 of 14 systems with spectroscopic 
data are probable binaries, consistent with the high binary 
frequency found for other massive stars in clusters and 
associations. 

\end{abstract}


\setcounter{footnote}{3}

\section{Introduction} 

The frequency of spectroscopic binaries (SBs) in star forming regions
has been a topic of interest for many years.  \citet{mas98} review the
fraction of binary systems found among O stars in clusters/associations,
the field, and runaway systems.  The results from their large survey
agree with previous studies \citep*[e.g.][]{gar80,lev91} and indicate 
that a large fraction of O stars in clusters and associations
are binaries while fewer field stars show multiplicity and very few
runaway stars have companions.  The values from \citet{mas98} are
broken down into spectroscopic and resolved binaries.  For O stars in
clusters and associations they give a total of 75\% as binaries; 61\%
of the binaries are spectroscopic and 41\% are resolved (with some obvious
overlap).  The spectroscopic binary group includes, however, 
a significant number of cases labeled as SB1? or SB2? (systems 
suspected of radial velocity variability or showing double lines),
so the fraction of confirmed spectroscopic binaries is then 34\% 
among O stars in clusters and associations.

We have conducted a radial velocity survey of O-type stars in the Cas OB6
stellar association in search of spectroscopic binary systems.  
The Cas~OB6 region is one of the largest complexes in the 
Perseus arm of the Galaxy and is located at a distance 
of about 2.3~kpc \citep*{gar92,mas95,fos03}.
The warm dust and ionized gas in this region extends 
over 150~pc along the arm and is dominated by the W3/W4/W5 chain of 
\ion{H}{2} regions and the HB3 supernova remnant \citep*{car00,ter03}.
The open cluster IC~1805 (OCL~352 = Melotte 15) is situated at the western 
end in the W4 region and contains 8 O-type stars \citep{mas95}.
The three brightest of these, HD~15558, 15570, and 15629, 
are probably the ionization source that powers the 
blowout of gas in the Galactic chimney \citep*{nor96,ter03} 
and the emission arc above W4 \citep*{rey01}.
The open cluster IC~1848 (OCL~364) is found in the eastern
part of the association in the W5 region.  
Its primary source of ionizing flux is the 
multiple O-star system HD~17505 \citep{sti01}. 
In addition to these two clusters, \citet{car00} find 
evidence of 19 other young clusters that are still embedded in 
their primordial clouds.  

Our knowledge of the binary star population among the young, 
massive stars of Cas~OB6 is quite limited at present. 
There are a number of radial velocity investigations of 
the sample but the overall small number of measurements has made 
the detection of binary stars difficult
\citep*{hay32,und67,ish70,con77,liu89,liu91,rau04}.  Spectroscopic
orbits are available for only three systems, HD~15558 \citep{gar81},
HD~16429 \citep{mcs03}, and BD$+60^\circ497$ \citep{rau04},
while a fourth star, BD$+60^\circ470$ = DN~Cas, is 
a known eclipsing binary \citep{fra74}. 
However, results for O-stars in other clusters and associations 
suggest that many more binaries may exist \citep{mas98,gar01}. 
It is important to determine the extent of the binary star 
population in Cas~OB6 in order to estimate accurately the 
ionizing flux of the key O-type stars and to assess the 
role of gravitational encounters with binary stars that 
might lead to ejection of runaway stars \citep*{gie86,hoo01}.

Here we present a reconnaissance of the radial velocity variations 
observed in 13 bright stars in Cas~OB6 that we obtained 
over a 5 night run with the KPNO 4~m Mayall telescope in 
2003 October.  We also present the first radial velocity study of the 
eclipsing binary system DN~Cas, which we combine with published photometry
to determine the component masses.  
We discuss our observations and reduction techniques in \S2.
The radial velocity measurements and the derivation of 
orbital elements are summarized in \S3.  
We use a tomography algorithm to separate the spectra of 
the binary star components, and we describe a comparison of 
the stellar spectra with model spectra in \S4 that we use 
to estimate the basic stellar parameters. 
We examine in \S5 the spectral energy distributions to determine
reddening, ratio of total-to-selective extinction, and angular 
size for each target, and we discuss the probable distance to Cas~OB6. 
Details about the individual objects follow in \S6. 
Finally we summarize in \S7 our derived fraction of multiple and 
single systems for Cas~OB6 in the context of past results.


\section{Spectroscopic Observations and Reductions} 

The blue spectra were obtained with the RC spectrograph
at the Mayall 4~m telescope at Kitt
Peak National Observatory from 2003 September 30 through 2003
October 4.  The spectra were made with the BL~380 grating 
(1200 grooves mm$^{-1}$, blazed at 4500 \AA ~in second order)
using a BG39 order sorting filter.  The detector was the T2KB CCD, 
a $2048\times2048$ device with 24 $\mu$m square pixels.   
The spectra recorded wavelengths between 4180 and 4940~\AA , 
although internal vignetting occurs for wavelengths
red-ward of about 4800~\AA .  The reciprocal dispersion of the spectra
is approximately 0.37 \AA ~pixel$^{-1}$, and the resolving power is 
$R=\lambda / \Delta\lambda = 5700$.  Exposure times were usually
a few minutes in length, and each stellar spectrum was preceded 
and followed by a He-Ne-Ar lamp exposure for careful 
wavelength calibration.  Most of the resulting spectra have a 
signal-to-noise ratio in excess of 100 pixel$^{-1}$. 
We usually obtained at least 
two spectra of each target on each of five consecutive 
nights.  We also made numerous bias and flat field 
frames each night.   A list of the names of targets 
appears in Table~1. 

\placetable{summary}  

We also obtained an additional 32 spectra of HD~17505A 
with the KPNO 0.9~m Coude Feed telescope in runs in 2000 October and 
2000 December \citep{boy05}.   The spectra were made with 
the long collimator, grating B (in second order with order
sorting filter OG550), camera 5, and a Ford $3072\times 1024$ 
CCD (F3KB) detector.
This arrangement produced a resolving power, 
$R=\lambda / \Delta\lambda = 9530$,
and covered a range of 829 \AA ~around H$\alpha$.  
Exposure times were 10 -- 15 minutes, and the resulting spectra
have a S/N = 200 in the continuum. 
We also observed the rapidly rotating A-type star, $\zeta$~Aql,
which we used for removal of atmospheric water vapor lines.
Each set of observations was accompanied by numerous bias, flat field,
and Th~Ar comparison lamp calibration frames.

The spectra were reduced using standard routines in IRAF\footnotemark
\footnotetext{IRAF is distributed by the National Optical Astronomical
Observatory, which is operated by the Association of Universities for
Research in Astronomy, Inc. (AURA), under cooperative agreement with the
National Science Foundation.} to create continuum rectified 
versions of each spectrum.  Then all the spectra of a given target
were transformed to a common heliocentric frame in $\log \lambda$ 
form and collected in a spectrum matrix of wavelength and time for 
subsequent analysis.   The rectification of the vignetted portion 
of the spectrum was generally problem-free, so we used the entire 
spectral range in the following analysis. 


\section{Radial Velocities and Orbital Elements} 

We found that the spectra of 10 of the 13 targets were 
composed of one set of spectral lines, and for these targets
we obtained radial velocities by finding the extremum point
of the cross-correlation of each spectrum with a spectrum 
template (omitting those spectral regions containing 
interstellar features).  We initially obtained velocities using 
the first spectrum as the template, and then we shifted 
each spectrum to the reference frame of the first spectrum
and co-added all the spectra to form a high S/N template. 
The final relative velocities were then found by cross-correlation 
with this high S/N template spectrum.  These relative velocities
were then transformed to absolute velocity by adding the 
velocity measured in the template spectrum by fitting 
parabolae to the lower portions of selected spectral lines.
Rest wavelengths were taken from the list of \citet{bol78}.
The final velocities for these 10 stars are listed in Table~2
(which appears only in the electronic version), 
and the mean velocity $<V_r>$ for each star is given in 
column 3 of Table~1. 
Note that the adopted absolute velocity of the template spectrum 
depends critically on what subset of lines is measured 
because of the presence of significant line-to-line velocity differences.  
These differences mainly result from line formation at 
differing heights in an expanding stellar atmosphere 
\citep{hut76,boh78,gie86}.  We generally selected high excitation
lines that form deep in the atmosphere where the outflow velocity 
is smallest (see discussion of the individual targets in \S6).
We list in column 4 of Table~1 $\sigma(l-l)$, the standard deviation of
the measured velocity among different lines in the template spectrum.  

\placetable{vrad} 

We used a version of the statistical $F$-test to determine if 
a star was a velocity variable object.  Since we usually obtained
at least two spectra every night of each object and since we do 
not expect large variations in velocity over the time span of a 
few hours, we estimated the average measurement error from the 
mean of the absolute value of the differences within a night 
divided by $\sqrt{2}$.   This spectrum-to-spectrum error estimate 
$\sigma(s-s)$ is given in column 5 of Table~1.  We then compared 
this to the overall standard deviation of the velocities $\sigma$
(column 6 of Table~1), and set the criterion for variability 
as $\sigma / \sigma(s-s) > 3.3$, which corresponds to the $<1\%$
probability that both variances are drawn from the same population
(for 9 degrees of freedom = 10 observations minus 1 mean value). 
We found that three of the ten single-lined targets were velocity 
variable according to this criterion (see the final column of 
Table~1).  The stars HD~17520 and BD$+60^\circ594$ are probably 
long period binary systems while the very luminous star HD~15570
exhibits variations related to atmospheric structure fluctuations 
(see the discussion in \S6). 

The spectra of the targets DN~Cas and BD$+60^\circ497$ both showed evidence 
of line-doubling in some observations, and the spectra of HD~17505A
displayed three components in the best separated cases. 
We were able to use a tomography algorithm to isolate the 
stationary line component in HD~17505A (\S6.5), 
and after removal of the stationary component the spectra
were treated in the same way as for the other double-lined systems. 
We measured velocities for these double-lined binaries 
in a two-step process \citep{gie02}.  
Preliminary radial velocities were found using the multiple
Gaussian fitting technique within the IRAF SPLOT routine.
Then preliminary orbital elements were determined 
that provided starting estimates of the velocities for 
both components in each spectrum.  
We then formed template spectra for both components by 
interpolating in the O-star synthetic spectrum grid of
\citet{lan03} for a temperature and gravity appropriate for 
each star, and these were convolved with broadening 
functions to account for rotational and instrumental broadening (\S4).  
Then we obtained velocities for specific spectral lines 
by a non-linear least squares fit of the observed profile 
with the two template spectra profiles co-added with 
a pre-defined monochromatic flux ratio (details are 
given for each case in \S6).   We took care to form 
an average velocity for each spectrum based on the same 
set of selected lines in order to avoid spurious variations
introduced by line-to-line velocity differences.  
The resulting velocities are given below in \S6 (Tables 4, 6, and 8).

We determined orbital elements for these three double-lined 
binaries using the non-linear, least-squares fitting routine of
\citet{mor74}.   The starting value for the orbital period was
obtained with the period search technique of \citet{sca82} as
modified by \citet{hor86} (or from photometry in the case of DN~Cas).
We first solved for the best fit period and then 
we formed separate elliptical solutions for both the primary 
and secondary components.  The derived eccentricity in the 
case of DN~Cas was not significantly different from zero
\citep{luc71}, but in the other two targets the eccentricity
was significant and the same within errors in the solutions 
for the primary and secondary components.  We formed average 
values of the eccentricity $e$, longitude of periastron $\omega$,
and time of periastron $T$ that were then fixed in solutions 
to obtain the semiamplitude $K$ and systemic velocity 
$\gamma$ for the primary and secondary components individually
(since differing atmospheric expansion rates may lead to 
different systemic velocities).  The final orbital elements 
are given in \S6 (Tables 5, 7, and 9).   


\section{Stellar Parameters} 

We co-added our spectra to form high S/N versions for use 
in spectral classification, temperature and gravity estimation, 
and measurement of the projected rotational velocity $V\sin i$.
We used a simple shift-and-add algorithm to form an average 
spectrum in the case of the single-lined targets.  However, 
we relied upon the Doppler tomography algorithm described by \citet{bag94}
to reconstruct the individual component spectra of the multiple
component systems.  The tomography algorithm is an iterative 
scheme that uses the derived orbital velocity shifts and 
monochromatic flux ratio to estimate the spectra of each component. 
We ran the algorithm for 50 iterations with a gain of 0.9, although 
the results are insensitive to both parameters.  The final, 
continuum rectified spectra of all our targets 
(except HD~17505; see \S6.5) and all their 
components are shown in Figures~1 and 2 in order of decreasing 
effective temperature. 

\begin{figure}
\begin{center}
\plotone{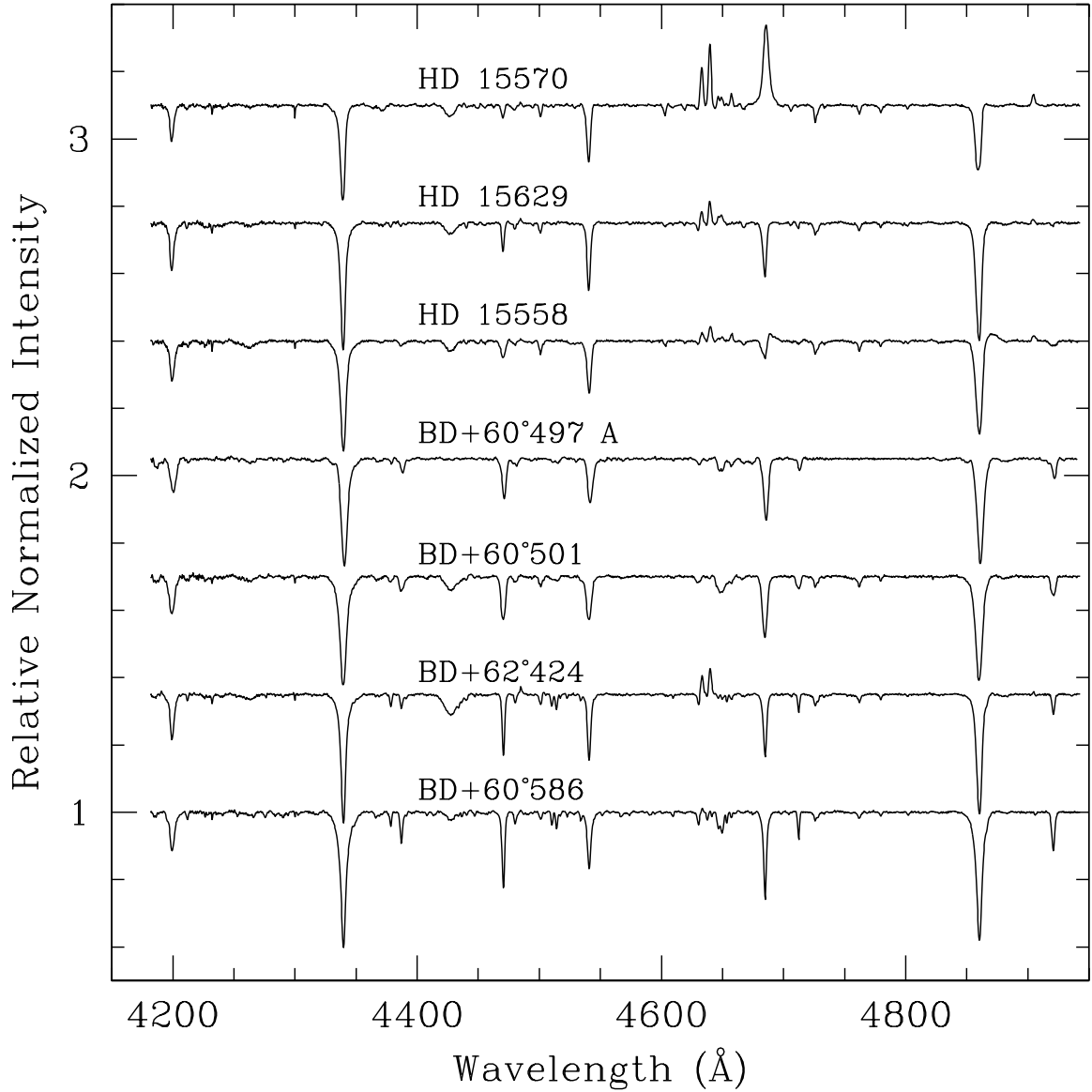}
\end{center}
\caption[The Observed Spectra sorted by Temperature] {The observed
mean spectra sorted by temperature.  Here we show the seven hottest
stars with the hottest at the top.  The spectra have
been continuum normalized and shifted vertically for clarity.
Note that interstellar features (such as the 4428 \AA ~diffuse
interstellar band) are cut out of the tomographic
reconstructions but are present in the other spectra. \label{specA}}
\end{figure}

\placefigure{specA}  

\begin{figure}
\begin{center}
\plotone{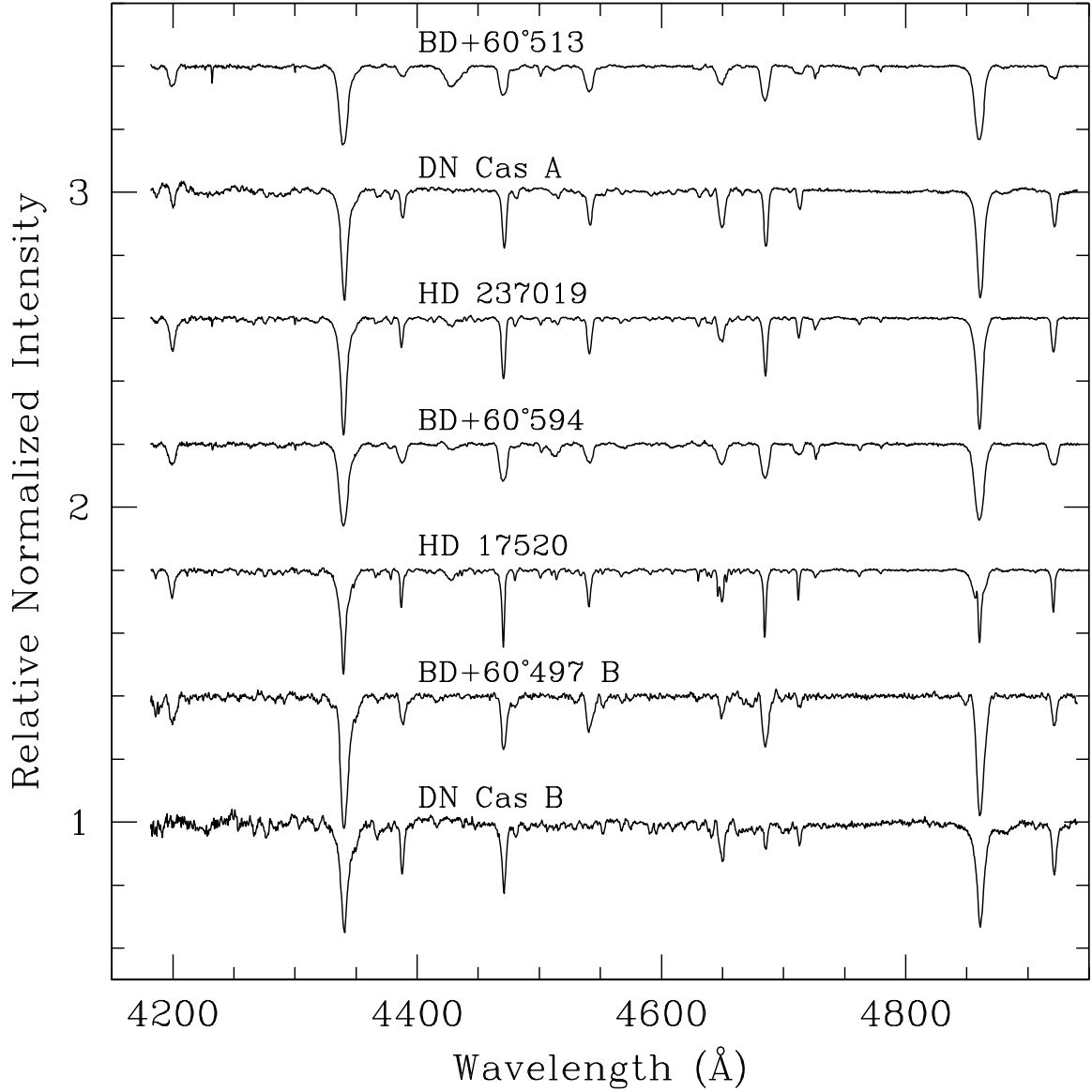}
\end{center}
\caption[The Observed Spectra sorted by Temperature] {The observed
spectra sorted by temperature.  Here we show the seven coolest
stars with the coolest at the bottom.  The spectra have
been continuum normalized and shifted vertically for clarity.
\label{specB}}
\end{figure}

\placefigure{specB}  

We estimated the spectral classification of each target (Table 1) by 
comparing their spectra with standard star spectra in the atlas 
of \citet{wal90}.    The corresponding stellar effective temperature
$T_{\rm eff}$ was determined by comparing the equivalent widths of 
various \ion{He}{2}/\ion{He}{1} line ratios with those measured 
in the synthetic spectral library for O-stars by \citet{lan03}. 
In most cases, we simply assumed a surface gravity of $\log g = 4.0$ 
as a suitable value for main sequence stars, but this approach 
was inadequate in several cases of evolved stars where the 
H Balmer line wings are narrower and indicative of a lower 
gravity.   We experimented with a number of model synthetic spectra
to best estimate the gravity in these cases.  This was accomplished
through comparing models by eye with the data in order
to find an appropriate match for surface gravity.  The most difficult 
spectrum to match was that of the very luminous star HD~15570, 
which displays clear evidence of the atmospheric expansion into a 
stellar wind that is not included in the static atmosphere 
formulation adopted by \citet{lan03}.  We estimate that our fitting 
errors are approximately $\pm 1000$~K in temperature and 
$\pm 0.3$ dex in $\log g$, and 
the results are listed in Table~1. 

We estimated the projected rotational velocity $V\sin i$ 
by comparing the observed \ion{He}{1} and \ion{He}{2} line widths 
with those found by convolving the model profiles from \citet{lan03} 
with instrumental and rotational broadening functions.  
The instrumental broadening of our spectra was represented
by a Gaussian with $\mathrm{FWHM}=52$ km s$^{-1}$.  
The rotational broadening function \citep{gra92} was 
computed assuming a spherical star with a linear limb-darkening
coefficient determined from \citet{wad85}.  A grid of
theoretical profiles was computed for a range of $V\sin i$,
and a $\chi^2$ minimization was used to determine the
best fit to the data.  The results and errors are listed in Table~1.
Note that this approach does not account for variations in 
the intensity profile across the disk, nor the non-spherical 
shape and gravity darkening found in rapid rotators 
\citep*{tow04}.  These shortcomings are most important for 
the two rapid rotators in our sample, BD$+60^\circ 513$ and
BD$+60^\circ594$, and we have probably underestimated the 
true rotational broadening in these cases. 


\section{Radii from Fits of the Spectral Energy Distributions} 

All of our targets are considered to be members of the 
Cas~OB6 association based upon their position in the sky 
and placement in the Hertzsprung-Russell diagram (HRD) 
\citep{vor85,gar92,wal02}. The proper motion study 
of IC~1805 made by \citet*{vas65}  confirms 
membership for five targets but curiously indicates 
a somewhat discrepant proper motion for HD~15558. 
We will assume that all the targets share a common distance, 
so their observed fluxes are directly related to their 
relative stellar radii.  Since we have determined temperature 
and multiplicity for each target, we have all parameters needed
to estimate their radii as a function of distance by comparing 
their observed and model flux distributions, and we can 
then check the distance by comparing the derived and model radii 
(as a function of temperature).  The angular diameter of the 
limb darkened disk $\theta_{LD}$ (in units of radians) is found by the 
inverse-square law:  
\begin{equation}
{ {f_\lambda ({\rm observed})}\over 
  {F_\lambda ({\rm emitted}) ~10^{-0.4 A_\lambda}} } =
  ( {R_\star / d} )^2 =
  {1\over 4} \theta_{LD}^2
\end{equation}
where the ratio of the observed and emitted fluxes 
(reduced by the effects of interstellar extinction $A_\lambda$)
depends on the square of the ratio of stellar radius $R_\star$ 
to distance $d$.   All our targets have a large reddening, 
so we need to fit the relation above over the entire 
observed spectrum in order to account reliably for the 
wavelength dependent shape of the extinction curve. 
We adopted the extinction curve law from \citet{fit99} 
that is a function of the reddening $E(B-V)$ and the ratio of 
total-to-selective extinction $R=A_V/E(B-V)$.   We then fit 
the observed flux distribution of each target to determine 
$E(B-V)$, $R$, and angular diameter $\theta_{LD}$. 

We compiled data on the observed fluxes from the UV to the near-IR.
There are low dispersion {\it International Ultraviolet Explorer (IUE)}
spectra available (SWP and LWR cameras) for all of the targets except 
DN~Cas, HD~17520, HD~237019, and BD$+60^\circ586$, and we formed binned 
versions of the UV spectral fluxes from these.  We found Johnson $UBV$ 
photometry for all the targets (plus Cousins $RI$ in a few cases), 
and these magnitudes were transformed to fluxes using the calibration 
of \citet*{col96}.  We also included Str\"{om}gren photometry for 
7 of the 13 targets using the calibration of \citet{gra98}.  
All of the stars are included in the {\it 2MASS All-Sky Catalog of Point Sources}
\citep{cut03}, and we converted the $JHK$ magnitudes to fluxes
using the calibration of \citet*{coh03}.  

The model fluxes were selected from the grid of \citet{lan03} 
based upon the values of $T_{\rm eff}$ and $\log g$ in Table~1.
These model fluxes are based upon plane-parallel, non-LTE,
line blanketed atmospheres that do not include the effects of
stellar wind mass loss.  However, the inclusion of winds generally
alters the spectral energy distributions only at wavelengths
much larger and much smaller than those considered here \citep*{mar05},
so wind effects are insignificant in this context.
We co-added model fluxes using the observed flux ratios 
for the spectroscopic binaries and using the observed magnitude 
difference for close visual binaries (details are given in next section 
for individual stars).   Note that in the two single-lined 
binary cases, HD~17520 and BD$+60^\circ594$, we simply assumed 
that the unseen companion contributes $5\%$ of the total flux
(a compromise between the expectation that the mass ratio is not 
too extreme and that brighter companions would have visible spectral 
features in our observations).  We have no spectral information 
for some of the companions, and in these cases we assumed that
the companion was a main sequence star with a temperature 
corresponding to the observed flux ratio (from the calibration 
in Table~4 of \citealt{mar05}).   We set a lower limit of 
$T_{\rm eff} = 27500$~K and $\log g=4.0$ for the faint companions, 
which corresponds to the cool boundary in the grid of \citet{lan03}. 
This simplification is acceptable since any detected companion 
is probably hot enough that the optical and near-IR regions 
correspond to the common Rayleigh-Jeans portion of the flux distribution.

We compared the observed and model flux distributions and 
made a grid search fit to find the values of $E(B-V)$ and $R$ that 
produced the lowest $\chi^2_\nu$ residuals (by weighting the 
observations approximately equally between the UV, optical, and 
near-IR bands).   The normalization term in the final fit yields
the angular diameter.   Our results are gathered 
in Table \ref{sedfits} which lists the star name, home cluster,
angular diameter (in micro-arcsec), derived radius (see below), 
$E(B-V)$, and $R$.  The acccompaning error estimates for 
$E(B-V)$ and $R$ are derived from the point in grid search where 
$\chi^2_\nu$ attains a value higher by 1 than normalized minimum value, 
and the errors in diameter are based upon the scatter in flux 
distribution fit and the errors propagated from uncertainty in 
$T_{\rm eff}$, $E(B-V)$, and $R$.  The model fits were generally 
excellent, but they were less satisfactory 
for the spectra of the three hottest stars, where it was difficult 
to find a consistent match of the far- and near-UV fluxes.  
We also found the $JHK$ fluxes were too large relative to the 
optical fluxes for HD~17520, which we suspect is due to an 
IR excess associated with the Be disk emission of the companion 
\citep{por03}, so the fit was based on the optical fluxes alone in this case. 

\placetable{sedfits} 

\begin{figure}
\begin{center}
\plotfiddle{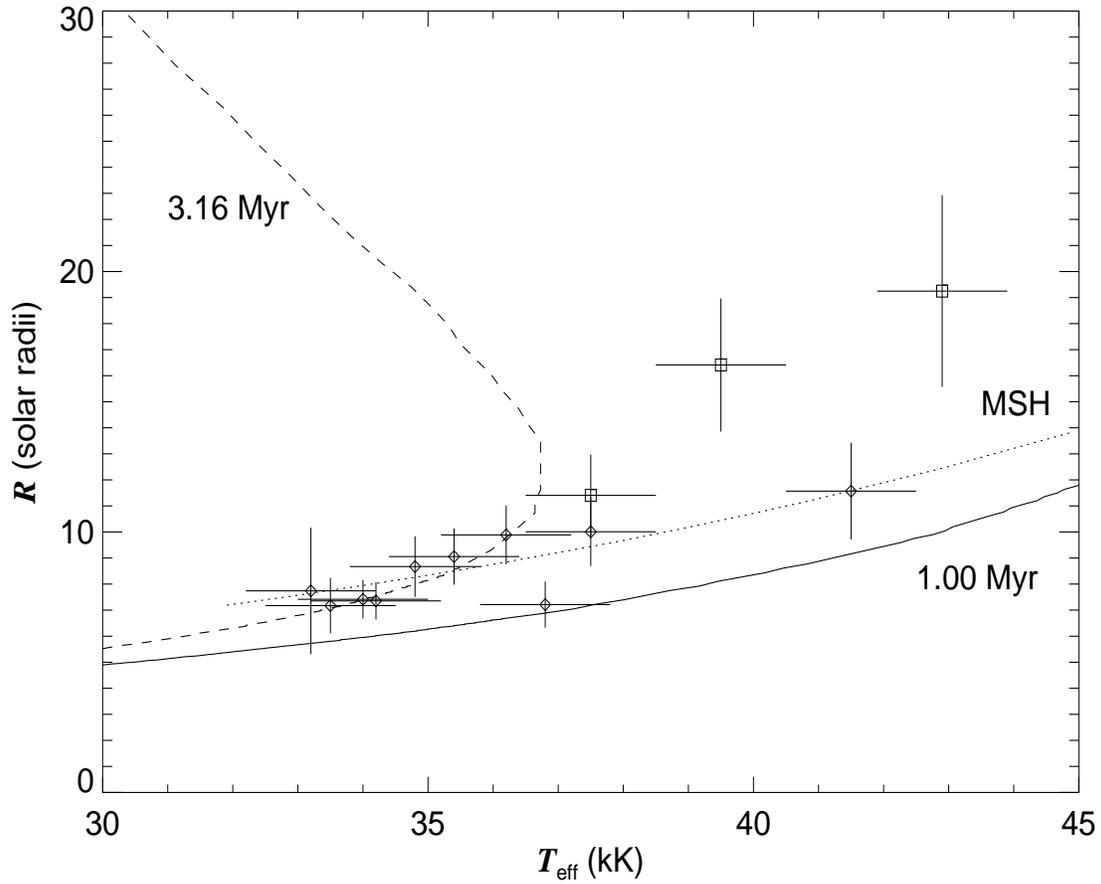}{100pt}{90}{360}{450}{30}{0}
\end{center}
\caption[Derived Stellar Radii and Temperatures] {The stellar 
radii (for an assumed distance of 1.9~kpc) plotted versus effective 
temperature ({\it diamonds} = V, {\it squares} = I, III luminosity classes).
The solid and dashed lines show the predicted radii for ages of 
1.00 and 3.16~Myr, respectively \citep{lej01}.
The dotted line indicates the relationship given for main sequence 
stars by \citet{mar05} (see their Table~4).  
\label{figradt}}
\end{figure}

The stellar radius is found by 
\begin{equation}
R_\star = 107.43~R_\odot ~\theta_{LD} ~d
\end{equation}
where the distance $d$ is given in kpc.   The distance estimates 
for IC~1805 and IC~1848 range from 1.9~kpc \citep{ish70} to 
2.4~kpc \citep{gue89} (with a scatter of $\pm 10\%$ among recent 
estimates).   We selected a distance of $d=1.886$~kpc from the
compendium of open cluster data by \citet{lok01} because 
this low end value appears to give the best match between the observed and 
theoretical stellar radii and provides good agreement with 
radii of similar stars in eclipsing binaries \citep{gie03}. 
The resulting radii are listed in Table \ref{sedfits} and are plotted 
against stellar temperature in Figure \ref{figradt}.  There is a clear 
and consistent trend for the hotter main sequence stars to be larger,  
and the observations match the predictions for main sequence stars 
from the atmosphere models of \citet{mar05} (shown as a dotted line 
in Fig.~\ref{figradt}).  We also show in Figure \ref{figradt} the 
($T_{\rm eff}, R_\star$) loci for stars with ages of 1.00 and 3.16~Myr 
from the evolutionary models of \citet{lej01}, which correspond to 
the age range for IC~1805 given by \citet{mas95}.   We see that most of the 
stars fall within the expected region in this diagram for this age range. 
However, the positions of several stars indicate that there is a 
real dispersion in the ages among this sample.  For example, 
BD$+60^\circ501$ appears to have a relatively small radius for its 
temperature, indicating it is the youngest object in our sample. 
The derived stellar radii are directly proportional to the assumed 
distance, and, for example, if we adopted the larger distance $d=2.34$~kpc 
advocated by \citet{mas95} then the stars would be approximately $24\%$ 
larger, a size that would be difficult to reconcile with the evolutionary
models of \citet{lej01}.  We show below that our derived radii 
for DN~Cas agree well with those determined independently from an analysis 
of the eclipsing light curve. 

\placefigure{figradt}  


\section{Results for Individual Objects} 

\subsection{Stars With Constant Radial Velocity} 

BD$+60^\circ501$.-- The two Balmer lines H$\gamma$ and
H$\beta$ have the lowest measured velocities in the 
template spectrum, so they were omitted from the average
(i.e., based only on the \ion{He}{1} and \ion{He}{2} lines).  
Our observed mean velocity of $-57.9\pm 1.1$ km~s$^{-1}$ 
is reasonably consistent with the mean from the 
work of \citet{rau04} of $-49.9\pm2.5$ km~s$^{-1}$
who also find the star to be a constant velocity object. 

HD 15558.-- There is evidence here for a velocity progression 
(lower velocity among low excitation lines), so we used 
only the lines \ion{He}{2} $\lambda\lambda 4199, 4541$ to 
set the radial velocity of the template spectrum.  
There are no obvious variations over the 5 night run, 
but the star may be a long period, single-lined binary 
\citep{gar81}.  According to the recent orbital 
elements from \citet{sti01}, the orbital period is 
$439.64\pm0.26$~d and the predicted velocity at the 
mean time of our observations is $-21.8$ km~s$^{-1}$ 
(near the peak of the radial velocity curve), which is 
27.7 km~s$^{-1}$ higher than our observed mean of  
$-49.5\pm2.9$ km~s$^{-1}$.  This discrepancy could be 
caused by differences in measurement techniques or errors in
the preliminary orbital solution of \citet{sti01}.  
A long-term study by M.\ De Becker and G.\ Rauw 
(2005, private communication) 
has recently confirmed the binary status of this star and will 
provide an improved orbital solution.  Therefore, we list
this star as ``constant (SB1)'' in Table \ref{summary} denoting
both a lack of short term line variability in our observations
as well as the recent confirmation of binarity.  There is a visual 
companion at a separation of $9\farcs9$ that is 2.6~mag 
fainter \citep{per97}.  

HD 15570.-- There is a clear velocity progression between 
spectral lines with the low excitation lines (H, \ion{He}{1}) 
appearing at a much lower velocity than the high excitation 
lines (\ion{He}{2}, \ion{N}{5}).  There is also a correlation 
with line width in the sense that stronger, broader lines 
have lower velocity.  We adopted a radial velocity for the 
template spectrum from the lines \ion{He}{2} $\lambda 4199$ 
and \ion{N}{5} $\lambda\lambda 4604, 4620$.  The night-to-night
scatter between velocities is small except on the last night 
when the measured velocity is $\approx 10$ km~s$^{-1}$ more 
positive than the mean from the other nights.  Curiously, 
the Of emission lines do not show any shift in velocity
on the final night.  We suspect that the change was related
to an atmospheric fluctuation rather than orbital motion. 
\citet{und90} also consider this star to be a velocity variable.
\citet*{deb05a} find no evidence of orbital motion and discuss
rotational modulation as a possible source of line variability.

HD 15629.-- The low excitation Balmer and \ion{He}{1} lines
have systematically low velocities, so the template mean was adopted 
from the lines \ion{He}{2} $\lambda\lambda 4199, 4541, 4686$.
We find no evidence of velocity variability.  
\citet{und90}, on the other hand, flag this star as 
possibly variable, but given the presence of line-to-line 
velocity differences, some of the velocity differences 
found by various authors are probably due to the line 
sample adopted (and not to orbital motion). 
\citet{deb05a} also find no evidence of orbital motion
for this star.

BD$+60^\circ513$.-- This star has a broad-lined spectrum, 
and we selected only the strongest lines to measure the 
template mean velocity (H$\gamma$, H$\beta$, 
\ion{He}{1} $\lambda\lambda 4471,4921$, 
\ion{He}{2} $\lambda\lambda 4541, 4686$).  
We confirm the lack of short term velocity variability 
reported by \citet{rau04}, and our mean velocity of 
$-58.6\pm 2.8$ km~s$^{-1}$ is probably consistent with 
their mean value, $-44.3\pm 10.5$ km~s$^{-1}$. 

BD$+62^\circ424$.-- The H$\beta$ line had the lowest 
velocity measured, so we omitted the H Balmer lines 
and weak lines in forming the template mean velocity. 
The star appears to be radial velocity constant over 
the run, and our mean velocity, $-34.3 \pm 1.7$ km~s$^{-1}$, 
is probably consistent with that from \citet{abt72} of 
$-43.5 \pm 6.3$ km~s$^{-1}$. 

HD 237019.-- All the strong lines were used to form the 
template mean velocity.  The star has a constant radial
velocity over the run.  Our mean velocity of 
$-43.4 \pm 1.1$ km~s$^{-1}$ is higher than the only 
other value in the literature of $-68$ km~s$^{-1}$ from 
\citet{sey41}. 

BD$+60^\circ586$.-- We omitted H$\beta$ and \ion{He}{1} $\lambda 4921$
from the template mean calculation, since both lines had 
significantly lower velocities.  The star was radial
velocity constant during the run.  Our mean velocity,
$-48.1\pm 1.0$ km~s$^{-1}$, agrees with that found by 
\citet{abt72}, $-42.3 \pm 7.1$ km~s$^{-1}$, but is lower
than the value reported by \citet{con77}, $-27.9\pm 1.2$ km~s$^{-1}$.
The star is a member of a visual multiple system (ADS~2194) with companions 
at separations of $7\farcs2$ ($\triangle V = 1.7$) and $13\farcs0$
($\triangle V = 4.7$). 

\subsection{Single-lined Spectroscopic Binaries} 

HD 17520.-- The H$\beta$ line has a blue-shifted emission component
which also appears weakly in H$\gamma$.  The Balmer lines and 
weak lines were omitted in the calculation of the 
template mean velocity.  The velocities show a progressive increase 
over the duration of the run, reaching values similar to
those observed by \citet{con77} and \citet{liu89}. 
\citet{wlt92} showed how H$\alpha$ emission developed 
between 1985 and 1991 leading to a spectral classification  
as a Be-type star (probably its status in 2003 as 
well since the emission is so apparent in H$\beta$).  
This star is also the central object in a visual multiple system (ADS~2165)
with a close companion at a separation of $0\farcs335$ 
and with a magnitude difference of 0.52~mag \citep{per97,wal02}. 
Our spectra show evidence of only one set of lines 
(for the O9~V component), and unfortunately, we do not know
if the Be emission originates in the O-star or its nearby 
companion (which could have broad shallow lines as  
observed in many Be stars).  We note that over the course 
of the run the H$\beta$ emission appeared to be almost 
stationary while the absorption lines shifted red-ward by 
$\approx 17$ km~s$^{-1}$.  The simplest interpretation is 
that the O-star component is a long-period, single-lined 
binary, while the Be emission originates in the fainter 
component of the close visual pair.  There is another star 
at a separation of $10\farcs8$ that is 3.2 mag fainter in $K$ 
\citep{cut03}. 

BD$+60^\circ594$.-- We used the strong lines to set the 
template mean velocity for the broad-lined spectrum of this
star.  The first measured velocity is close to the value 
obtained by \citet{con77}, $-50.5\pm5.6$ km~s$^{-1}$, 
but the subsequent velocities decline monotonically by 
$\approx 49$ km~s$^{-1}$ over the next 4 days.  This 
suggests that the star is a single-lined binary with a
period of $\approx 20$~d and a semiamplitude of 
$\approx 50$ km~s$^{-1}$. 

\subsection{The Double-Lined Eclipsing System DN~Cas} 

DN Cas has been known as an eclipsing binary for some time
and was classified as an Algol type binary with a period
of 1.155479 d by \citet{hof47}.
\citet{hil56} classified the spectrum as O8~V.  
A later photometric study by \citet{fra74} used $UBV$ 
photoelectric photometry to refine the
orbital period to 2.310955 d, twice that of the original
value.  Unfortunately, the comparison star used by \citet{fra74}
turned out to be itself an eclipsing binary, compromising their
final photometry.  \citet{dav80} made another photometric study, 
and based upon the curvature near the maxima of the light curve 
he reclassified DN~Cas as a $\beta$~Lyrae type system. 
Interestingly no spectroscopic observations of DN~Cas have
appeared in the literature, though \citet{dav80} claimed that such
observations were underway.  Additionally, the only photometry
that has been published, other than in phase-folded figures,
is the compromised data of \citet{fra74}.  This overlooked 
opportunity to obtain masses for this massive binary is 
rectified here, where we present the first full orbital solution 
based on our spectroscopy and photometry from the literature.

We determined velocities for both components using non-linear 
least-squares fits of two templates with temperatures and 
rotational velocities given in Table~\ref{summary} and with a monochromatic 
flux ratio in the blue of $F_2/F_1=0.41\pm 0.10$.  
Our final radial velocities are given in Table \ref{dncasrv} 
and the spectroscopic orbital elements are given in 
Table \ref{dncasorb}.  The period was adopted from the eclipse timing
observations collected by \citet*{kre01}.  Here we adopt the 
time of minimum light (primary superior conjunction) to set the 
zero-point of orbital phase and epoch $T_{\rm min}$, and 
our derived epoch agrees with the current photometric 
epoch of minimum light from \citet{kre01} of HJD 2,452,501.928 $\pm 0.007$. 
Eccentric fits offered no significant improvement over 
circular fits \citep{luc71}, so circular elements are given in Table~5. 
The radial velocities and orbital solutions are shown in Figure~4.

\placetable{dncasrv}  

\placetable{dncasorb}  

\begin{figure}
\begin{center}
\plotone{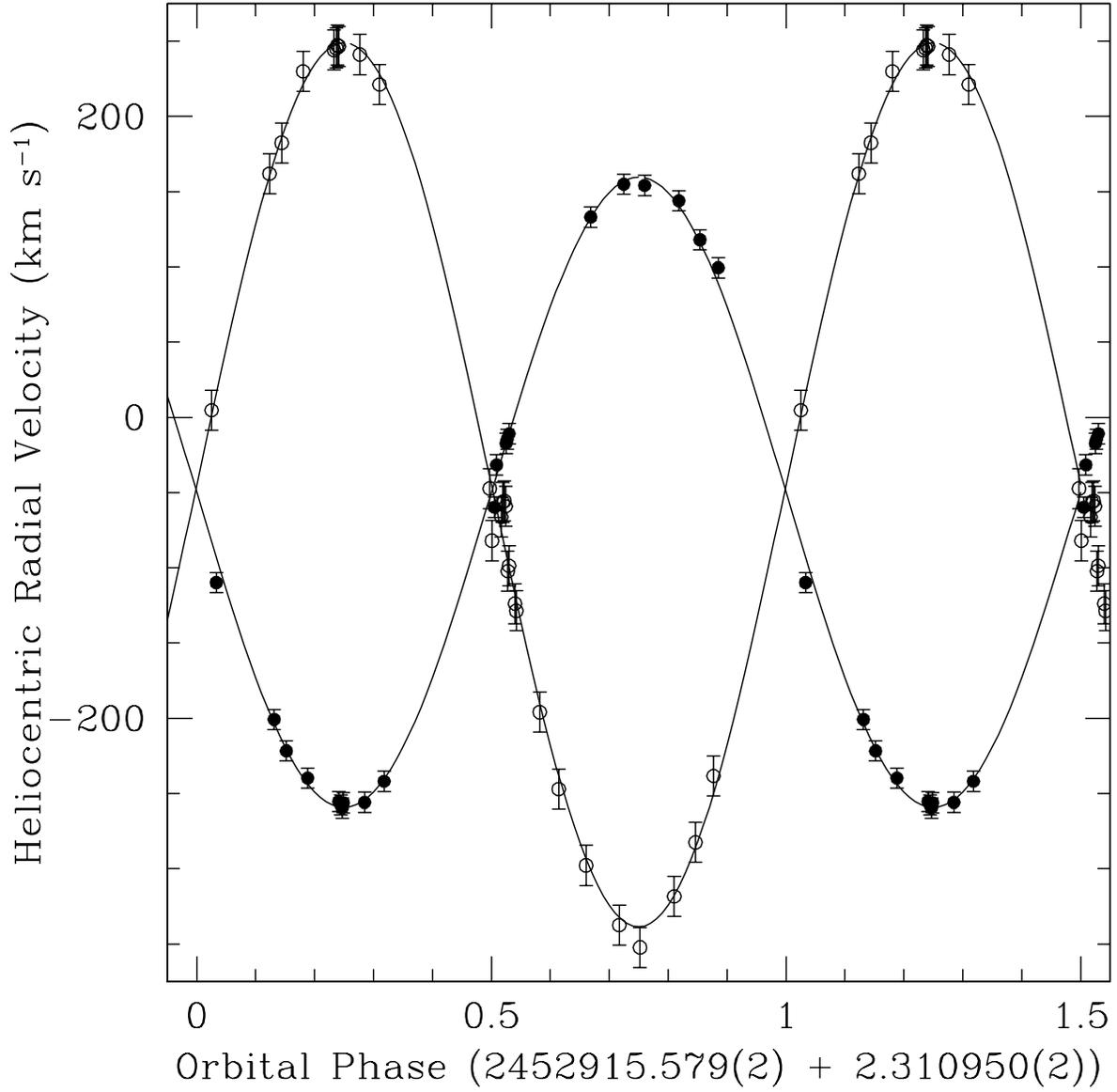}
\end{center}
\caption[DN Cas Phase-folded Radial Velocity Curve] {The DN Cas
radial velocity data folded on the given ephemeris and plotted with the
calculated orbital velocity curves.
\label{dncasph}}
\end{figure}

\placefigure{dncasph}  

The only published photometric data of DN~Cas are those from 
\citet{fra74} who made $UBV$ photoelectric observations between 
1971 October and 1972 December.  Unfortunately, their comparison
star, HD~14817, is itself an eclipsing binary as mentioned above.
However, the light curve of HD~14817 from {\it Hipparcos} \citep{per97}
is flat outside of eclipse, so we can retain the results 
from \citet{fra74} that are based on outside-of-eclipse times
for HD~14817.  These measurements are plotted against orbital 
phase in Figure~5, and because there remain significant phase 
gaps in this restricted set, our light curve fitting results 
should be considered preliminary.

\begin{figure}
\begin{center}
\plotone{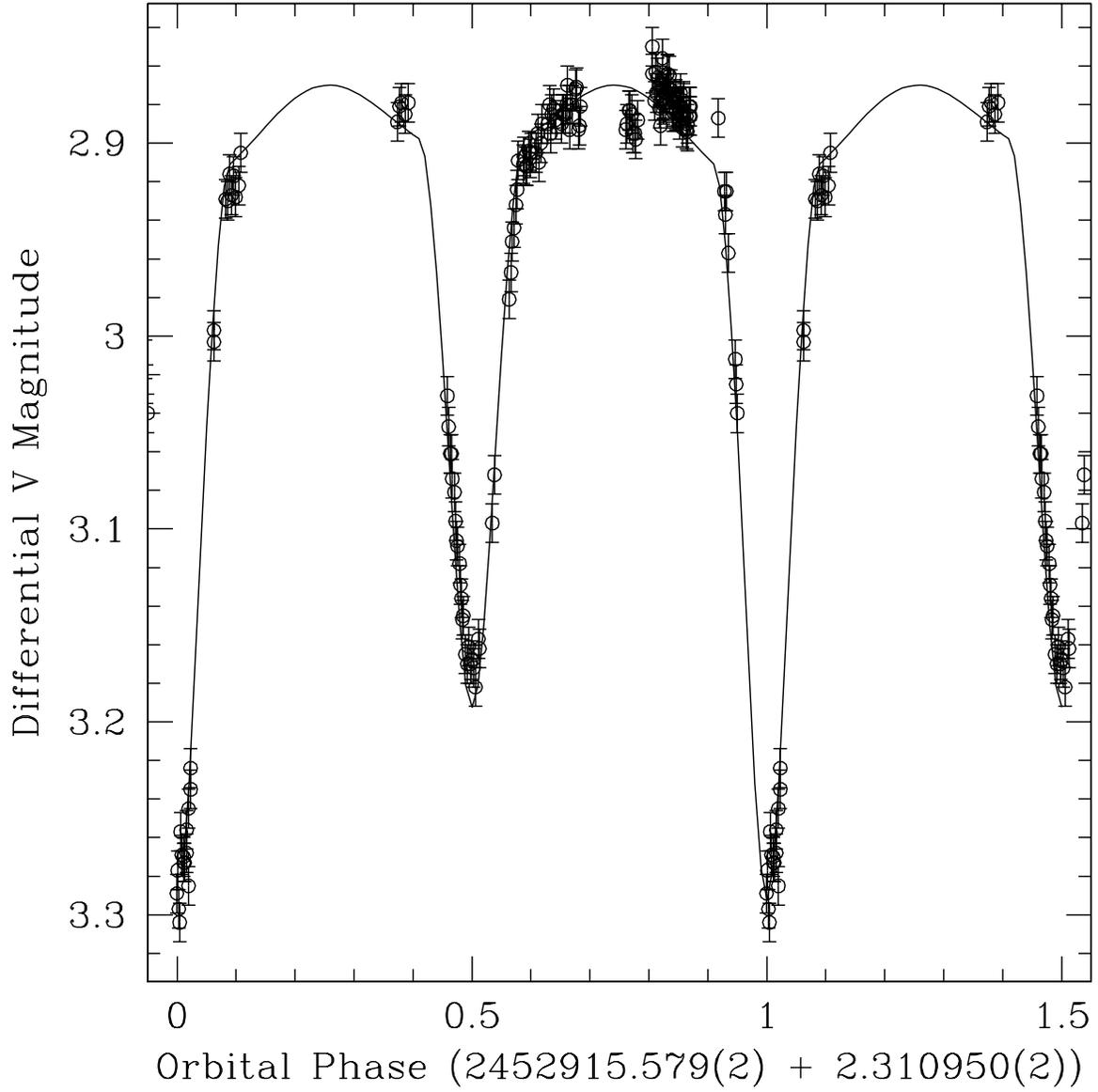}
\end{center}
\caption[DN Cas Phase-folded Light Curve] {The DN Cas $V$-band
photometry data folded on the given ephemeris and plotted with the
Wilson-Devinney code light curve.
\label{dncaslc}}
\end{figure}

\placefigure{dncaslc}  

We calculated a model light curve using the Wilson-Devinney code
\citep{wil71,wil90}.  Most of the model parameters are set 
by the spectroscopic orbital solution (Table~\ref{dncasorb}), 
the adopted temperatures and gravities from the spectra (Table~\ref{summary}),
the associated limb darkening parameters \citep{van93}, 
and the monochromatic flux ratio (which sets the ratio of 
the mean stellar radii).  In addition, we set the bolometric albedo 
and gravity darkening exponents to 1.0, which are 
valid approximations for stars with fully radiative envelopes.
We then fit the observed $V$-band observations by varying 
the inclination $i$ and the stellar radii.
The ratio of stellar radii was set using the observed 
monochromatic flux ratio and model surface flux ratio 
(set according to the temperatures and gravities in Table~1)
$$R_2/R_1 = [ (f_\lambda ({\rm observed})_2 / f_\lambda ({\rm observed})_1) 
 \times (F_\lambda ({\rm emitted})_1 / F_\lambda ({\rm emitted})_2) ]^{0.5} = 0.72 \pm 0.09.$$
If the stars are rotating synchronously, then the ratio of 
projected rotational velocities also yields the ratio of radii as
$$R_2/R_1 = (V\sin i)_2 / (V\sin i)_1 = 0.72 \pm0.15$$
and the good agreement between these two estimates suggests 
that synchronous rotation has been attained in this close binary.  
We found that the radii obtained from a fit of the spectral energy distribution
and distance of 1.9~kpc (Table \ref{sedfits}) led to an excellent fit
of the light curve for an orbital inclination of 
$i=79\fdg2 \pm 1\fdg1$ (Fig.~5), and the derived 
stellar masses are $M_1=19.2\pm 0.8~M_\odot$
for the O8~V primary and $M_2=13.9\pm 0.5~M_\odot$ for
the B0.2~V secondary (Table~5).  The mass of the primary 
is consistent with its spectral type for the calibration 
given  by \citet{mar05}, and the mass of the secondary agrees
well with the mass for its spectral type according to the relations given by \citet*{han97}.
Both stars are well within their respective Roche lobes. 

There is a companion to DN~Cas in the 2MASS catalogue at a separation of
$5\farcs3$ and at a position angle of $89^\circ$.  However, inspection of
the 2MASS images at this position suggests that this object is
probably not a real star since its point spread function looks very sharp
and unlike those of nearby stars.  Furthermore, we see no evidence
of such a tertiary in our observations.  Since the near east-west alignment
of the putative object and DN~Cas is the same as the slit orientation for our 
KPNO 4~m spectra, the tertiary should appear in our spectra with a 
separation from DN~Cas of 7.7 pixels in the spatial dimension.
The average 2MASS magnitude difference is 2.13 mag (for $H,K$),
so that the flux ratio of tertiary to binary is 0.14 in the near-IR. 
However, our blue spectra show no sign of a companion brighter than
$\triangle m = 5$, so if this companion is real, it is probably a 
very red foreground star that can be ignored in our analysis.   
We found that the introduction of a third light contribution 
of $14\%$ in the $V$-band light curve results in an increase in the
inclination by $3^\circ$, but since the actual flux contribution 
of a tertiary is probably much smaller than this, any changes 
in our results due to the presence of a tertiary are probably 
comparable to or less than the errors quoted in Table~5. 

\subsection{The Double-Lined System BD+60$^\circ$497} 

The spectra of BD+$60^\circ$497 show clear evidence that this star
is a double-lined spectroscopic binary.  
In their recent paper, \citet{rau04} give the first orbital solution
for BD+$60^\circ$497, and they find an orbital period of 3.96 days.
We measured velocities using the scheme outlined in \S3 by matching 
template synthetic profiles for models with temperatures and gravities 
given in Table~1, together with an estimated flux ratio of 
$F_2/F_1=0.35\pm 0.08$.
We note, however, that in some of the lines (for example, \ion{He}{1}
$\lambda\lambda 4387, 4471$) we found that the approaching, 
blue-shifted component (primary and secondary) often appeared 
to be weaker than expected based upon this fixed flux ratio 
(\citealt*{rau04} also observed a similar variation).  
We doubt that this variation has a significant effect on 
our measurements, which appear in Table~6.

\placetable{60d497rv}  

We combined our velocities with those of \citet{rau04} in 
making the orbital solution.  A number of alias periods remain 
viable, but we obtained the best fit with a period close to that
obtained by \citet{rau04} (see Table~7).  However, unlike 
\citet{rau04}, we found that a significant eccentricity is 
present, and we made full elliptical fits for both components. 
Otherwise our results are in reasonable agreement with 
their solution (compared in Table~7). 
The velocities and orbital fit are shown folded on the derived ephemeris in
Figure~6.   Assuming masses typical of their spectral classes 
\citep{mar05}, the system probably has an inclination
of $i\approx 47^\circ$, so we do not expect eclipses to occur.  

\placetable{60d497rv}  

\begin{figure}
\begin{center}
\plotone{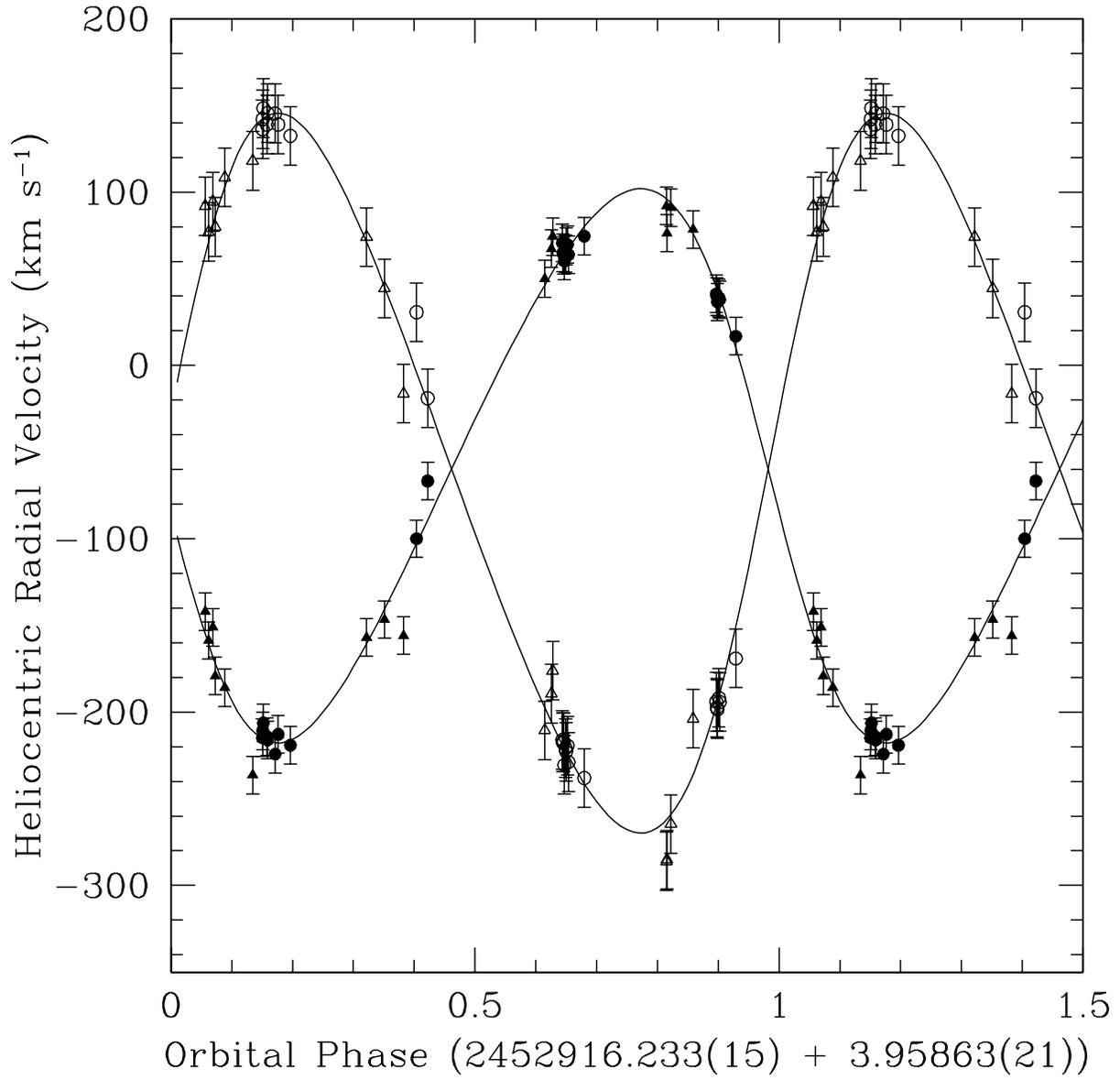}
\end{center}
\caption[BD+$60^\circ$497 Phase-folded Radial Velocity Curve] {Our 
radial velocity data ({\it circles}) with the data from \citet{rau04}
({\it triangles}) for BD$+60^\circ497$.  The data is folded on the
given ephemeris and plotted with the calculated model 
({\it line}).  The solid points denote radial velocities 
of the primary and the open points represent secondary star 
velocities. \label{60d497ph}}
\end{figure}

\placefigure{60d497ph}  

\subsection{The Multiple Star System HD~17505} 

The star system HD~17505 dominates the center of the cluster IC~1848, 
and there are seven visual companions noted by \citet{mas98} 
at distances ranging from $2\farcs1$ to $124\farcs0$. 
Multiplicity in the central object HD~17505A has
been discussed by \citet{con71}, \citet{wal73}, \citet{how97},
and \citet{wal02} and it is known to contain at least 
three stars since the single {\it International Ultraviolet Explorer}
spectrum shows evidence of three Doppler shifted components 
\citep{sti01}.  \citet{wal02} pointed out that the multiplicity 
of the star is the reason it appears so bright in the HRD.
We obtained 10 spectra of HD~17505A over the
five night duration of the KPNO 4~m run, and we found that
the lines were strongly blended in all but a few spectra 
corresponding to the maximum Doppler separation of the binary motion.
However, we also obtained an additional 32 red spectra of the target 
in 2000 with the KPNO 0.9~m Coude Feed telescope (\S3), and these
turned out to be particularly helpful in finding the orbital elements. 

Initial inspection of these higher dispersion red spectra showed that 
all three components were cleanly resolved in the \ion{He}{1} profiles 
observed near orbital quadrature phases and indicated an orbital 
period of approximately 8~days.   We measured radial velocities by  
first making unconstrained three-Gaussian fits of the least 
blended feature, \ion{He}{1} $\lambda 7065$, in the quadrature phase spectra.
We obtained a preliminary orbital solution from this subset of spectra
for the two moving components, and then we used this solution to 
estimate the Doppler shifts for all the Coude Feed observations.  
Then we made three-component Gaussian fits of \ion{He}{1} $\lambda 7065$
in all the red spectra using these starting estimates for the Doppler 
shifts and constraining the widths to be the mean values found from 
the fits of the quadrature phase spectra.  A new orbital solution was 
obtained, and the solution was used to make a three component reconstruction 
of the individual component spectra using the Doppler tomography algorithm 
\citep{bag94}.  We subtracted the reconstructed spectrum of the stationary 
component from each spectrum and renormalized them to end up with a 
set corresponding to the spectra of the double-lined binary alone. 
We then measured the radial velocities using the non-linear, 
least-squares method by matching template spectra to the profiles 
of H$\alpha$, \ion{He}{1} $\lambda 6678$/\ion{He}{2} $\lambda 6683$, 
and \ion{He}{1} $\lambda 7065$.  The template spectra for the two 
components were taken from their respective tomographic reconstructions
($F_2/F_1=1.0\pm 0.1$) rather than using the model profiles from 
\citet{lan03}, because we were concerned that some of these red 
features are not well represented by the synthetic models. 
This approach worked well except in one conjunction phase 
observation (from HJD~2,451,892.743) where the lines 
were too badly blended for reliable measurement. 
Our final radial velocities from the red spectra for the close 
binary are collected in Table \ref{17505rv}.

\placetable{17505rv} 

The estimate for the orbital period from the red spectra alone was 
$P= 8.572\pm0.003$~d, and the error was sufficiently small that we could 
determine the cycle number and approximate orbital phase corresponding 
to the times of the KPNO 4~m observations in 2003 October.  
We measured the line width of the \ion{He}{1} $\lambda 4471$ feature 
in these blue spectra to find that maximum velocity separation occurred 
at HJD~2,452,915.469 $\pm 0.12$, and we compared this to the time 
of the same orbital phase for the 2000 Coude Feed observations to 
arrive at a revised period estimate of $P= 8.5710\pm0.0008$~d. 
The final orbital solution is
given in Table~9 and the radial velocity curve appears in Figure~7. 
The predictions from this fit 
for the time of the {\it IUE} observation are $+128$ and $-184$ km~s$^{-1}$,
in reasonable agreement with the observed velocities of
$+142$ and $-165$ km~s$^{-1}$ \citep{sti01}, but we did not include the 
{\it IUE} measurements in the final solutions because of possible systematic 
differences between velocities measured in the UV and optical. 

\placetable{17505orb} 

\begin{figure}
\begin{center}
\plotone{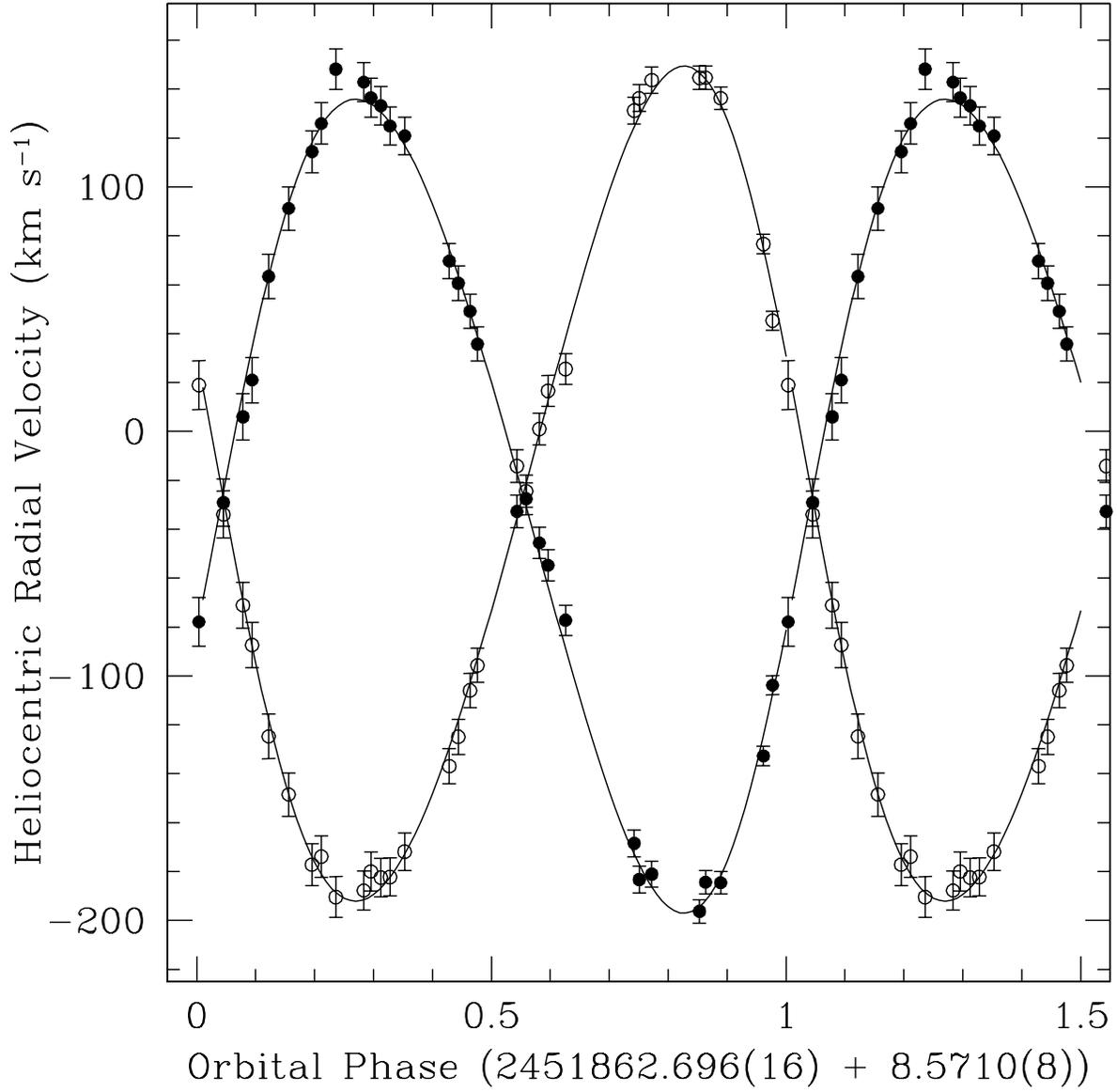}
\end{center}
\caption[HD~17505Aa Phase-folded Radial Velocity Curve] {The
HD~17505Aa radial velocity data folded on the given ephemeris,
plotted with the calculated radial velocity curve.
\label{hd17505ph}}
\end{figure}

\placefigure{hd17505ph} 

\begin{figure}
\begin{center}
\epsscale{0.9}
\plotone{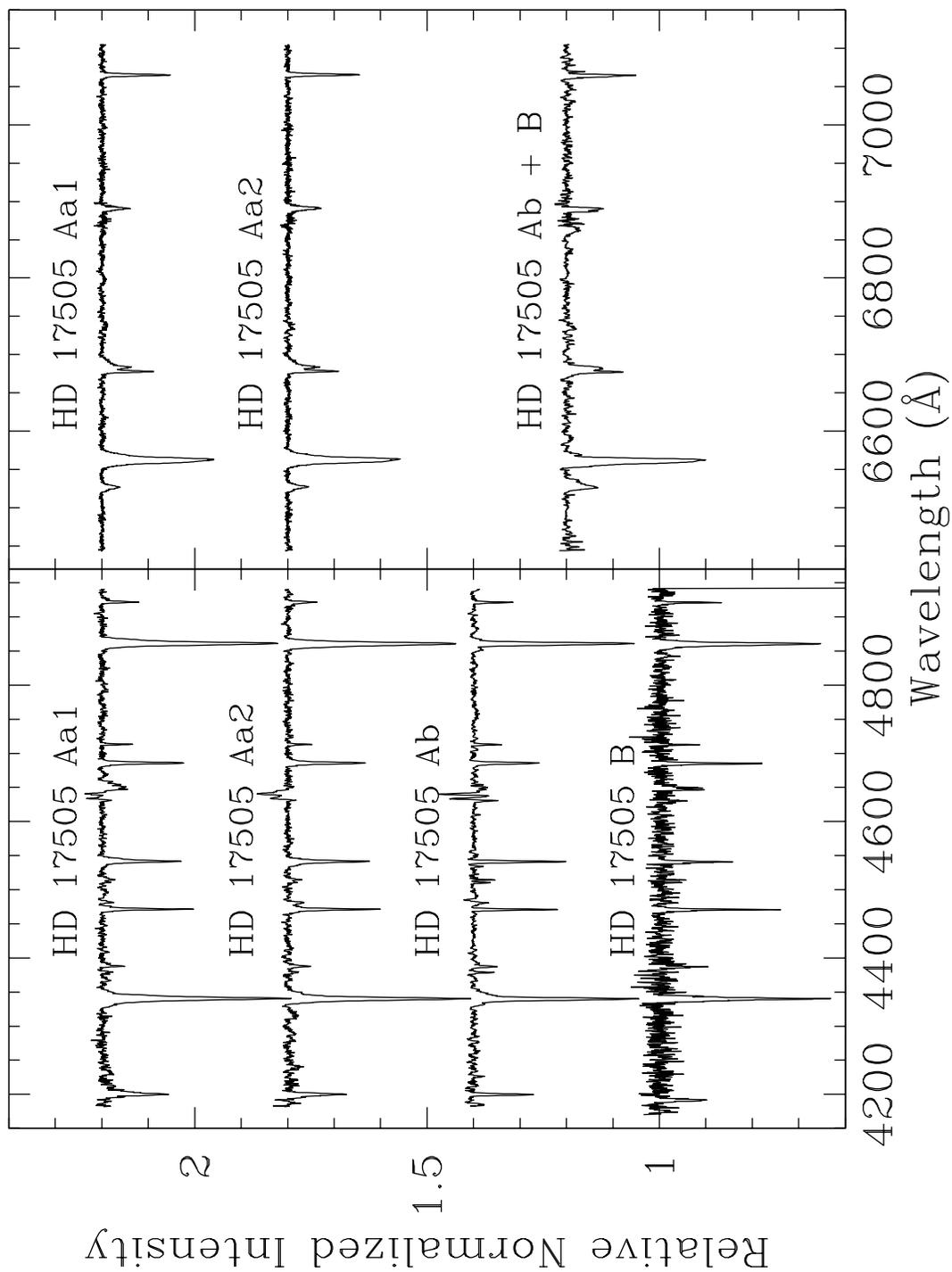}
\caption[Blue and Red Spectra of HD~17505] {Spectra for the
components of HD~17505 in both the blue and red wavelength regions.
\label{17505spec}}
\end{center}
\end{figure}

\begin{figure}
\begin{center}
\plotone{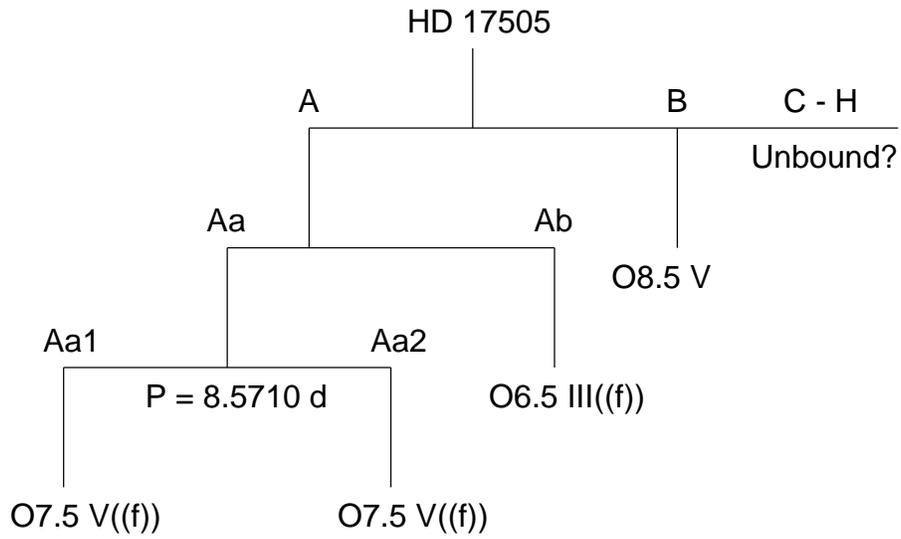}
\end{center}
\caption[Mobile Diagram of HD~17505A] {Diagram showing the
hierarchy of the HD~17505 system.
\label{17505mob}}
\end{figure}

We made tomographic reconstructions of both the blue and red spectra
using the orbital solution in Table \ref{17505orb} and assuming flux
contributions of 30\%, 30\%, and 40\% for Aa1, Aa2, and Ab, respectively
(based upon the relative equivalent widths of the \ion{He}{1} $\lambda 6678$
$+$ \ion{He}{2} $\lambda 6683$ feature). 
Both the blue and red spectral reconstructions are plotted in
Figure \ref{17505spec}.   We determined radial velocities 
of the stationary component Ab of $-30 \pm 3$ and $-38 \pm 5$
from the blue and red reconstructed spectra, respectively.
Once again, the optical velocities are systematically more negative than 
the UV determined value for Ab of $-13.5$ km~s$^{-1}$ \citep{sti01}.
Our results are summarized in a schematic representation of the
system hierarchy in Figure \ref{17505mob}.  Note that the 
Coude Feed spectra were made with a $2\farcs5$ wide slit in an approximately north-south 
orientation, so the B component (at a separation of $2\farcs15$ and a position 
angle of $93^\circ$ with a magnitude difference of $\triangle H_p = 1.66$ mag; 
\citealt{per97}) may be slightly mixed in with the light from 
the Ab component for the red reconstructed spectrum.  However, the 4~m spectra
were made with an east-west slit orientation, so it was possible to separate 
the A and B components spatially along the slit.  A single extracted spectrum
of the B component is shown in Figure \ref{17505spec} along with the tomographic
reconstructions of the Aa1, Aa2, and Ab components. The spectrum of the B
component is clearly different and indicates a cooler effective temperature
than the  stars of component A, and, thus, the spectroscopic triple we detect
is clearly  related to A alone.   The hottest and brightest of all the
components is Ab, and its spectral type agrees with that obtained by
\citet{wal72} for the entire system. 

\placefigure{17505spec}  

\placefigure{17505mob}   

The stars of the Aa1-Aa2 central binary are far enough apart that no eclipses 
should be seen, and indeed none are observed \citep{per97}.  For example, 
if we adopt the masses and radii associated with O-type main sequence stars
from \citet{mar05}, then the expected system inclination should be
$i=65^\circ$ (based upon the $M\sin^3 i$ values in Table~\ref{17505orb})
while eclipses would be observed 
only if $i>73^\circ$ (based upon the semimajor axis in Table~\ref{17505orb}). 
The orbital period of B around the triple A can be estimated as $\approx 27,000$~y
if we take the projected angular separation as the semimajor axis and the 
distance is 1.9~kpc.  The orbital period of the Aa-Ab central triple is 
probably much less.  The system was unresolved in the speckle survey of 
\citet{mas98}, and if we take their resolution limit of 35~mas as the projected 
semimajor axis then the orbital period should be $<61$~y.  The target is 
clearly worthy of continued spectroscopic and interferometric 
observation to search for evidence of such orbital motion.   


\section{Summary} 

We have monitored thirteen O stars in the Cas~OB6 association
in search of multiple star systems.  Five of the thirteen were
found to have radial velocity variations related to 
orbital motion.  Of these five, we have determined orbital 
solutions for three double-lined systems: 
the eclipsing binary DN~Cas, BD$+60^\circ497$, and the 
spectroscopic triple HD~17505A.  The remaining two binaries,
BD$+60^\circ594$ and HD~17520, are single-lined binaries with 
periods longer than 5 days.
In addition to these five, one of our targets, HD~15558, is 
probably a long period binary \citep{gar81,sti01}.  
One other bright O-star in Cas~OB6 is HD~16429, which 
was found by \citet{mcs03} to be a spectroscopic triple. 
If we include all of these objects in the known sample of observed
O-stars in the Cas~OB6 association, then we find that 7 of the 14, or
$50 \%$, are spectroscopic binaries.  This value is well within
the range found by \citet{mas98} for spectroscopic binaries among
stars in clusters and associations.

We note that two systems, HD~17505 and DN~Cas, are of special 
interest and would benefit from follow-up observations.  
We are planning to obtain new three-color photometry of the
eclipsing binary DN~Cas in order to make a better fit
of the light curve and to determine an accurate mass of 
the O8~V primary star. 
The multiple system HD~17505 is a fascinating system with at least
four O stars (with a total mass close to $100~M_\odot$) 
which are apparently gravitationally bound.  High
resolution studies via interferometry to separate the Aa and Ab components
would produce a more accurate kinematic model for the system
and would help us better understand stellar interactions in
these systems and their role in the formation and ejection
of massive stars.


\acknowledgements

We are grateful to the director and staff of KPNO 
for their help in making these observations possible. 
We also thank Gregor Rauw and Michael De Becker 
for sharing their results with us in advance of publication.  
Financial support was provided by the National Science
Foundation through grant AST$-$0205297 (DRG).
Institutional support has been provided from the GSU College
of Arts and Sciences and from the Research Program Enhancement
fund of the Board of Regents of the University System of Georgia,
administered through the GSU Office of the Vice President
for Research.  MVM is supported by an NSF Astronomy and Astrophysics
Postdoctoral Fellowship under award AST$-$0401460.
This publication makes use of data products from the Two Micron All
Sky Survey, which is a joint project of the University of Massachusetts
and the Infrared Processing and Analysis Center/California Institute of
Technology, funded by the National Aeronautics and Space Administration
and the National Science Foundation.
This research has also made use of the Washington Double Star Catalog 
maintained at the U.S.\ Naval Observatory.





\begin{center}
\begin{deluxetable}{lccccccccc}
\tabletypesize{\scriptsize}
\tablewidth{0pc}
\tablecolumns{10}
\tablecaption{Target Object Summary\label{summary}}
\tablehead{
\colhead {Object}        & 
\colhead {Spectral}      & 
\colhead {$<V_r>$\tablenotemark{a}}       &
\colhead {$\sigma(l-l)$} & 
\colhead {$\sigma(s-s)$} &
\colhead {$\sigma$}      & 
\colhead {$T_{\rm eff}$} & 
\colhead {log $g$}       &
\colhead {$V\sin i$}     & 
\colhead {Binary}       \\
\colhead {Name}          & 
\colhead {Class.}        & 
\colhead {(km s$^{-1}$)} &
\colhead {(km s$^{-1}$)} & 
\colhead {(km s$^{-1}$)} & 
\colhead {(km s$^{-1}$)} & 
\colhead {(kK)}          & 
\colhead {(cm s$^{-2}$)} &
\colhead {(km s$^{-1}$)} & 
\colhead {Status}        }
\startdata 
DN Cas A          & O8 V        &   $-52$&\nodata& 4.9   & 149.3     & 34.2 & 4.0 &   $158\pm20$   & SB2  \\
DN Cas B          & B0.2 V      &   $-38$&\nodata& 5.3   & 207.3     & 29.7 & 4.0 &   $113\pm18$   & SB2  \\
BD$+60^\circ497$~A& O6 V        &   $-43$&\nodata& 2.2   & 126.8     & 37.5 & 4.0 &   $183\pm9$\phn& SB2  \\
BD$+60^\circ497$~B& O8 V        &   $-81$&\nodata& 3.6   & 166.7     & 33.0 & 4.0 &   $171\pm19$   & SB2  \\
BD$+60^\circ501$  & O6.5 V      &   $-58$&  9.4  & 0.5   &\phn\phn1.1& 36.8 & 4.0 &   $173\pm10$   & constant  \\
HD~15558          & O5 III(f)   &   $-49$&  5.3  & 1.8   &\phn\phn2.9& 39.5 & 3.7 &   $195\pm11$   & constant (SB1) \\
HD~15570          & O4 If+      &   $-55$&  0.9  & 0.8   &\phn\phn4.2& 42.9 & 3.5 &   $123\pm10$   & atmospheric var.\\
HD~15629          & O5 V((f))   &   $-60$&  8.1  & 1.2   &\phn\phn2.2& 41.5 & 4.0 &   $106\pm6$\phn& constant  \\
BD$+60^\circ513$  & O7.5 Vn     &   $-59$&  6.7  & 1.5   &\phn\phn2.8& 34.8 & 4.0 &   $259\pm5$\phn& constant  \\
BD$+62^\circ424$  &O6.5 V((f))  &   $-34$&  3.7  & 0.8   &\phn\phn1.7& 36.2 & 3.8 &   $103\pm12$   & constant  \\
HD~17505 Aa1      &O7.5 V((f))  &\phn$-9$&\nodata& 8.6   & 115.3     & 36.3 & 4.0 &\phn$90\pm10$   & SB2 \\
HD~17505 Aa2      &O7.5 V((f))  &   $-44$&\nodata& 7.2   & 118.2     & 36.5 & 4.0 &   $120\pm10$   & SB2 \\
HD~17505 Ab       &O6.5 III((f))&   $-38$&  8.2  &\nodata&\nodata    & 37.5 & 3.8 &\phn$80\pm15$   & constant \\
HD~17520          & O9 V        &   $-54$&  2.7  & 0.7   &\phn\phn6.0& 33.2 & 4.0 &\phn$80\pm12$   & SB1 + Be  \\
HD~237019         & O8 V        &   $-43$&  8.1  & 0.7   &\phn\phn1.1& 34.0 & 4.0 &   $128\pm5$\phn& constant  \\
BD$+60^\circ586$  & O7.5 V      &   $-48$&  4.7  & 1.0   &\phn\phn1.0& 35.4 & 4.0 &\phn$95\pm5$\phn& constant  \\
BD$+60^\circ594$  & O8.5 Vn     &   $-86$&  8.3  & 1.0   &   \phn17.7& 33.5 & 4.0 &   $285\pm16$   & SB1       \\
\enddata
\tablenotetext{a}{Systemic velocities for the binary stars are given in Tables \ref{dncasorb}, 
   \ref{60d497orb}, and \ref{17505orb}.}
\end{deluxetable}
\end{center}

\clearpage


\begin{center}
\begin{deluxetable}{lcr}
\tabletypesize{\scriptsize}
\tablewidth{0pc}
\tablecolumns{3}
\tablecaption{Radial Velocity Measurements for Constant Velocity and SB1
Stars\label{vrad}}
\tablehead{
\colhead {Star}  & \colhead {Date} & \colhead {$V_r$} \\
\colhead {Name}  & \colhead {(HJD-2,450,000)} & \colhead {(km s$^{-1}$)} }
\startdata 
BD$+60^\circ501$& 2912.891 & $-56.6$ \\
BD$+60^\circ501$& 2912.968 & $-56.4$ \\
BD$+60^\circ501$& 2913.888 & $-57.2$ \\
BD$+60^\circ501$& 2913.965 & $-57.9$ \\
BD$+60^\circ501$& 2914.890 & $-57.4$ \\
\enddata
\tablecomments{Table \ref{vrad} is published in its entirety in the electronic edition of
the Astrophysical Journal.  A portion is shown here for guidance regarding its form and content.}
\end{deluxetable}
\end{center}

\clearpage


\begin{deluxetable}{lccccc}
\tablewidth{0pc}
\tablecaption{Spectral Energy Distribution Fits\label{sedfits}}
\tablehead{
\colhead{Object} &
\colhead{Cluster} &
\colhead{$\theta_{LD}$} &
\colhead{$R_\star$\tablenotemark{a}} &
\colhead{$E(B-V)$} &
\colhead{$R$} \\
\colhead{Name} &
\colhead{Membership} &
\colhead{($10^{-6}$ arcsec)} &
\colhead{($R_\odot$)} &
\colhead{(mag)} &
\colhead{($A_V/E(B-V)$)} }
\startdata
DN Cas A           \dotfill & IC 1805 & 36 (1)\phn & \phn 7.4 (7)\phn &  0.97 (1) &  3.13 (1)\phn\phn \\
DN Cas B           \dotfill & IC 1805 & 26 (3)\phn & \phn 5.3 (8)\phn &  0.97 (1) &  3.13 (1)\phn\phn \\
BD$+60^\circ497$ A \dotfill & IC 1805 & 49 (4)\phn &     10.0 (13)    &  0.89 (2) &  2.94 (6)\phn\phn \\
BD$+60^\circ497$ B \dotfill & IC 1805 & 32 (5)\phn & \phn 6.5 (11)    &  0.89 (2) &  2.94 (6)\phn\phn \\
BD$+60^\circ501$   \dotfill & IC 1805 & 36 (3)\phn & \phn 7.2 (9)\phn &  0.76 (2) &  3.12 (5)\phn\phn \\
HD 15558 A         \dotfill & IC 1805 & 81 (10)    &     16.4 (26)    &  0.84 (3) &  3.07 (9)\phn\phn \\
HD 15558 B         \dotfill & IC 1805 & 28 (4)\phn & \phn 5.6 (9)\phn &  0.84 (3) &  3.07 (9)\phn\phn \\
HD 15570           \dotfill & IC 1805 & 95 (16)    &     19.3 (37)    &  1.01 (2) &  3.13 (7)\phn\phn \\
HD 15629           \dotfill & IC 1805 & 57 (7)\phn &     11.6 (19)    &  0.76 (3) &  3.09 (12)\phn    \\
BD$+60^\circ513$   \dotfill & IC 1805 & 43 (4)\phn & \phn 8.7 (12)    &  0.83 (3) &  3.00 (5)\phn\phn \\
BD$+62^\circ424$   \dotfill & IC 1805 & 49 (3)\phn & \phn 9.9 (11)    &  0.75 (2) &  3.02 (4)\phn\phn \\
HD 17505 Ab        \dotfill & IC 1848 & 56 (6)\phn &     11.4 (16)    &  0.72 (3) &  2.83 (9)\phn\phn \\
HD 17505 Aa1       \dotfill & IC 1848 & 49 (8)\phn &     10.0 (18)    &  0.72 (3) &  2.83 (9)\phn\phn \\
HD 17505 Aa2       \dotfill & IC 1848 & 49 (8)\phn &     10.0 (18)    &  0.72 (3) &  2.83 (9)\phn\phn \\
HD 17505 B         \dotfill & IC 1848 & 44 (6)\phn & \phn 8.9 (15)    &  0.72 (3) &  2.83 (9)\phn\phn \\
HD 17520 Aa        \dotfill & IC 1848 & 38 (11)    & \phn 7.7 (24)    &  0.57 (5) &  3.10 (120)       \\
HD 17520 Ab        \dotfill & IC 1848 & 14 (4)\phn & \phn 2.8 (9)\phn &  0.57 (5) &  3.10 (120)       \\
HD 17520 B         \dotfill & IC 1848 & 34 (10)    & \phn 7.0 (22)    &  0.57 (5) &  3.10 (120)       \\
HD 17520 C         \dotfill & IC 1848 & 14 (4)\phn & \phn 2.8 (9)\phn &  0.57 (5) &  3.10 (120)       \\
HD 237019          \dotfill & IC 1848 & 37 (1)\phn & \phn 7.4 (7)\phn &  0.77 (1) &  3.20 (6)\phn\phn \\
BD$+60^\circ586$ A \dotfill & IC 1848 & 45 (3)\phn & \phn 9.1 (11)    &  0.58 (3) &  3.27 (32)\phn    \\
BD$+60^\circ586$ B \dotfill & IC 1848 & 25 (2)\phn & \phn 5.1 (7)\phn &  0.58 (3) &  3.27 (32)\phn    \\
BD$+60^\circ594$ A \dotfill & IC 1848 & 35 (4)\phn & \phn 7.2 (11)    &  0.68 (2) &  2.88 (10)\phn    \\
BD$+60^\circ594$ B \dotfill & IC 1848 & 10 (2)\phn & \phn 2.0 (5)\phn &  0.68 (2) &  2.88 (10)\phn    \\
\enddata
\tablenotetext{a}{Radius derived for a distance of 1.886 kpc.}
\end{deluxetable}

\clearpage


\begin{center}
\begin{deluxetable}{cccccc}
\tabletypesize{\scriptsize}
\tablewidth{0pc}
\tablecolumns{6}
\tablecaption{Radial Velocity Measurements for DN Cas\label{dncasrv}}
\tablehead{
\colhead {Date}  & \colhead {Orbital} & \colhead {$V_1$} &
\colhead {$(O-C)_1$}  & \colhead {$V_2$} & \colhead {$(O-C)_2$} \\
\colhead {(HJD-2,450,000)}  & \colhead {Phase} & \colhead {(km s$^{-1}$)} &
\colhead {(km s$^{-1}$)}  & \colhead {(km s$^{-1}$)} & \colhead {(km s$^{-1}$)} }
\startdata 
2912.849  &  0.814 & 
\phs         $ 143.9$ &\phn     $  -1.4$ &  
             $-318.3$ &\phn     $  -7.1$  \\
2912.932  &  0.850 & 
\phs         $ 118.0$ &\phn     $  -3.8$ &  
             $-282.5$ &\phn     $  -3.8$  \\
2913.003  &  0.881 & 
\phn\phs     $  99.4$ &\phn\phs $   4.6$ &  
             $-238.3$ &\phn\phs $   2.9$  \\
2913.825  &  0.237 & 
             $-255.4$ &\phn\phs $   2.7$ &  
\phs         $ 244.1$ &\phn     $  -4.2$  \\
2913.834  &  0.241 & 
             $-257.6$ &\phn\phs $   0.8$ &  
\phs         $ 245.4$ &\phn     $  -3.4$  \\
2913.839  &  0.243 & 
             $-259.7$ &\phn     $  -1.1$ &  
\phs         $ 247.4$ &\phn     $  -1.6$  \\
2913.844  &  0.245 & 
             $-256.1$ &\phn\phs $   2.6$ &  
\phs         $ 246.6$ &\phn     $  -2.5$  \\
2913.927  &  0.281 & 
             $-255.9$ &\phn     $  -1.0$ &  
\phs         $ 241.1$ &\phn     $  -2.7$  \\
2914.004  &  0.314 & 
             $-241.9$ &\phn\phs $   0.1$ &  
\phs         $ 221.2$ &\phn     $  -4.7$  \\
2914.814  &  0.665 & 
\phs         $ 133.1$ &\phn\phs $   0.4$ &  
             $-297.8$ &\phn     $  -4.0$  \\
2914.943  &  0.721 & 
\phs         $ 154.9$ &\phn     $  -3.7$ &  
             $-337.5$ &\phn     $  -7.7$  \\
2915.025  &  0.756 & 
\phs         $ 154.1$ &\phn     $  -8.0$ &  
             $-352.3$ &         $ -17.8$  \\
2915.656  &  0.029 & 
             $-109.8$ &         $ -23.1$ &  
\phn\phn\phs $   4.7$ &\phn     $  -5.9$  \\
2915.883  &  0.127 & 
             $-200.8$ &\phn     $  -1.5$ &  
\phs         $ 161.9$ &\phn     $  -4.8$  \\
2915.931  &  0.148 & 
             $-221.7$ &\phn     $  -4.6$ &  
\phs         $ 182.5$ &\phn     $  -9.0$  \\
2916.015  &  0.184 & 
             $-239.8$ &\phn\phs $   1.4$ &  
\phs         $ 229.9$ &\phn\phs $   5.1$  \\
2916.747  &  0.501 & 
\phn         $ -59.8$ &         $ -13.1$ &  
\phn         $ -47.3$ &\phn     $  -2.4$  \\
2916.756  &  0.505 & 
\phn         $ -31.6$ &\phs     $  10.1$ &  
\phn         $ -82.0$ &         $ -30.1$  \\
2916.792  &  0.521 & 
\phn         $ -17.3$ &\phn\phs $   3.4$ &  
\phn         $ -66.2$ &\phs     $  14.7$  \\
2916.797  &  0.523 & 
\phn         $ -14.6$ &\phn\phs $   3.6$ &  
\phn         $ -56.4$ &\phs     $  28.2$  \\
2916.805  &  0.526 & 
\phn         $ -10.9$ &\phn\phs $   3.0$ &  
\phn         $ -55.6$ &\phs     $  34.9$  \\
2916.809  &  0.528 & 
\phn         $ -18.8$ &\phn     $  -7.5$ &  
\phn         $ -59.1$ &\phs     $  35.1$  \\
2916.818  &  0.532 & 
\phn\phn\phs $   3.7$ &\phs     $  10.3$ &  
             $-102.2$ &\phn     $  -1.6$  \\
2916.822  &  0.534 & 
\phn\phn\phs $   8.7$ &\phs     $  12.6$ &  
\phn         $ -98.6$ &\phn\phs $   5.7$  \\
2916.846  &  0.544 & 
\phn\phs     $  17.7$ &\phn\phs $   8.4$ &  
             $-123.8$ &\phn     $  -1.1$  \\
2916.851  &  0.546 & 
\phn\phs     $  14.4$ &\phn\phs $   2.5$ &  
             $-128.6$ &\phn     $  -2.3$  \\
2916.943  &  0.586 & 
\phn\phs     $  64.2$ &\phn\phs $   4.3$ &  
             $-195.9$ &\phn     $  -3.2$  \\
2917.018  &  0.618 & 
\phn\phs     $  92.2$ &\phn     $  -1.9$ &  
             $-247.1$ &\phn     $  -6.8$  \\
\enddata
\end{deluxetable}
\end{center}

\clearpage


\begin{deluxetable}{lr}
\tablewidth{0pc}
\tablecaption{Orbital Elements for DN~Cas \label{dncasorb}}
\tablehead{
\colhead{Element} &
\colhead{Value} }
\startdata
$P$~(days)                  \dotfill & 2.310950 (2) \\
$T_{\rm min}$ (HJD-2,450,000) \dotfill & 2915.579 (2) \\
$e$                         \dotfill & 0 \\
$\omega$ ($^\circ$)         \dotfill & \nodata \\
$K_1$ (km s$^{-1}$)         \dotfill & 210.5 (20) \\
$K_2$ (km s$^{-1}$)         \dotfill & 292.0 (38) \\
$M_2/M_1$                   \dotfill & 0.721 (12) \\
$\gamma$ (km s$^{-1}$)      \dotfill & $-48.3$ (14) \\
$\gamma$ (km s$^{-1}$)      \dotfill & $-42.7$ (27) \\
$i$ ($^\circ$)              \dotfill & 79.2 (11) \\
$M_1$ ($M_\odot$)           \dotfill &  19.2 (8) \\
$M_2$ ($M_\odot$)           \dotfill &  13.9 (5) \\
$R_1$ ($R_\odot$)           \dotfill &  7.4 (6) \\
$R_2$ ($R_\odot$)           \dotfill &  5.4 (8) \\
$R_1/R_1({\rm lobe})$       \dotfill &  0.78 (6)  \\
$R_2/R_2({\rm lobe})$       \dotfill &  0.66 (10)  \\
$a$ ($R_\odot$)             \dotfill & 23.3 (2) \\
r.m.s.$_1$ (km s$^{-1}$)    \dotfill & 7.4 \\
r.m.s.$_2$ (km s$^{-1}$)    \dotfill & 14.1 \\
\enddata
\tablecomments{Numbers in parentheses give the error in the last digit
quoted.}
\end{deluxetable}

\clearpage


\begin{center}
\begin{deluxetable}{cccccc}
\tabletypesize{\scriptsize}
\tablewidth{0pc}
\tablecolumns{6}
\tablecaption{Radial Velocity Measurements for BD$+60^\circ497$\label{60d497rv}}
\tablehead{
\colhead {Date}  & \colhead {Orbital} & \colhead {$V_1$} &
\colhead {$(O-C)_1$}  & \colhead {$V_2$} & \colhead {$(O-C)_2$} \\
\colhead {(HJD-2,450,000)}  & \colhead {Phase} & \colhead {(km s$^{-1}$)} &
\colhead {(km s$^{-1}$)}  & \colhead {(km s$^{-1}$)} & \colhead {(km s$^{-1}$)} }
\startdata 
2912.871 & 0.151 &      $-214.8$ & \phs\phn  0.6 & \phs   136.3 & \phn  $-6.1$ \\
2912.953 & 0.171 &      $-224.3$ & \phn   $-6.5$ & \phs   145.4 & \phn  $-0.1$ \\
2913.874 & 0.404 &      $-100.0$ & \phs\phn  2.7 & \phs\phn 30.7 & \phs    34.7 \\
2913.946 & 0.422 & \phn  $-66.7$ & \phs     22.4 & \phn $-18.9$ & \phs\phn 2.7 \\
2914.823 & 0.644 & \phs\phn 70.7 & \phs\phn  8.0 &     $-216.0$ & \phs\phn 2.7 \\
2914.826 & 0.644 & \phs\phn 65.0 & \phs\phn  2.0 &     $-217.6$ & \phs\phn 1.7 \\
2914.832 & 0.646 & \phs\phn 63.6 & \phn   $-0.3$ &     $-217.5$ & \phs\phn 2.9 \\
2914.836 & 0.647 & \phs\phn 60.1 & \phn   $-4.3$ &     $-230.5$ & \phn  $-9.5$ \\
2914.844 & 0.649 & \phs\phn 64.5 & \phn   $-0.9$ &     $-220.8$ & \phs\phn 1.4 \\
2914.848 & 0.650 & \phs\phn 68.6 & \phs\phn  2.6 &     $-222.9$ & \phs\phn 0.2 \\
2914.857 & 0.652 & \phs\phn 69.8 & \phs\phn  2.6 &     $-219.4$ & \phs\phn 5.1 \\
2914.862 & 0.654 & \phs\phn 63.9 & \phn   $-3.8$ &     $-228.9$ & \phn  $-3.6$ \\
2914.965 & 0.680 & \phs\phn 74.6 & \phn   $-5.4$ &     $-238.0$ & \phs\phn 3.1 \\
2915.825 & 0.897 & \phs\phn 41.3 & \phn   $-3.9$ &     $-194.0$ & \phs\phn 1.9 \\
2915.829 & 0.898 & \phs\phn 39.8 & \phn   $-4.4$ &     $-197.5$ & \phn  $-2.8$ \\
2915.833 & 0.899 & \phs\phn 36.6 & \phn   $-6.6$ &     $-198.4$ & \phn  $-5.0$ \\
2915.842 & 0.901 & \phs\phn 38.9 & \phn   $-2.0$ &     $-191.8$ & \phn  $-1.3$ \\
2915.846 & 0.902 & \phs\phn 38.0 & \phn   $-2.0$ &     $-194.2$ & \phn  $-5.0$ \\
2915.952 & 0.929 & \phs\phn 16.9 & \phs\phn  6.6 &     $-169.0$ &      $-18.3$ \\
2916.831 & 0.151 &      $-210.9$ & \phs\phn  4.6 &   \phs 142.0 & \phn  $-0.5$ \\
2916.835 & 0.152 &      $-206.3$ & \phs\phn  9.3 &   \phs 148.4 & \phs\phn 5.7 \\
2916.859 & 0.158 &      $-214.1$ & \phs\phn  2.5 &   \phs 139.0 & \phn  $-5.0$ \\
2916.863 & 0.159 &      $-216.2$ & \phs\phn  0.6 &   \phs 145.6 & \phs\phn 1.4 \\
2916.932 & 0.176 &      $-212.9$ & \phs\phn  5.0 &   \phs 139.0 & \phn  $-6.7$ \\
2917.011 & 0.196 &      $-219.1$ & \phn   $-2.8$ &   \phs 132.5 &      $-11.1$ \\
\enddata
\end{deluxetable}
\end{center}

\clearpage


\begin{deluxetable}{lcc}
\tablewidth{0pc}
\tablecaption{Orbital Elements for BD$+60^\circ497$ \label{60d497orb}}
\tablehead{
\colhead{Element} &
\colhead{Rauw \& De Becker (2004)} & 
\colhead{This paper} }
\startdata
$P$~(days)                  \dotfill & 3.96 (9)      & 3.95863 (21) \\
$T$ (HJD-2,450,000)         \dotfill & \nodata       & 2916.233 (15) \\
$T_0$ (HJD-2,450,000)       \dotfill & 2935.98 (15)  & \nodata  \\
$e$                         \dotfill & 0             & 0.156 (19) \\
$\omega$ ($^\circ$)         \dotfill & \nodata       & 100 (11) \\
$K_1$ (km s$^{-1}$)         \dotfill & 172 (12)      & 159.9 (22) \\
$K_2$ (km s$^{-1}$)         \dotfill & 221 (15)      & 207.7 (29) \\
$M_2/M_1$                   \dotfill & 0.78 (8)      & 0.770 (15) \\
$\gamma_1$ (km s$^{-1}$)    \dotfill & $-54$ (9)     & $-53.8$ (17) \\
$\gamma_2$ (km s$^{-1}$)    \dotfill & $-69$ (12)    & $-67.5$ (23) \\
$M_1$ sin$^{3}i$ ($M_\odot$)\dotfill & 13.9 (25)     & 11.1 (5) \\
$M_2$ sin$^{3}i$ ($M_\odot$)\dotfill & 10.9 (20)     & 8.6 (4) \\
$a$ sin $i$ ($R_\odot$)     \dotfill & 30.7 (19)     & 28.39 (28) \\
r.m.s.$_1$ (km s$^{-1}$)    \dotfill & \nodata       & 10.9 \\
r.m.s.$_2$ (km s$^{-1}$)    \dotfill & \nodata       & 14.4 \\
\enddata
\tablecomments{Numbers in parentheses give the error in the last digit
quoted.}
\end{deluxetable}

\clearpage


\begin{deluxetable}{lccccc}
\tabletypesize{\scriptsize}
\tablewidth{0pt}
\tablecaption{Radial Velocity Measurements for HD 17505Aa\label{17505rv}}
\tablehead{
\colhead{HJD}             &
\colhead{Orbital}         &
\colhead{$V_1$}           &
\colhead{$(O-C)_1$}       &
\colhead{$V_2$}           &
\colhead{$(O-C)_2$}       \\
\colhead{(-2,450,000)}    &
\colhead{Phase}           &
\colhead{(km s$^{-1}$)}   &
\colhead{(km s$^{-1}$)}   &
\colhead{(km s$^{-1}$)}   &
\colhead{(km s$^{-1}$)}   }
\scriptsize
\startdata
1817.887 &  0.772 &        $-$181.1 & \phn\phs5.3 &       \phs143.7 &
\phn\phs5.2 \\
1818.890 &  0.889 &        $-$184.6 &  \phn$-$2.8 &       \phs136.3 &
\phn\phs2.6 \\
1819.873 &  0.004 &     \phn$-$77.9 &  \phn$-$1.3 &    \phn\phs18.9 & 
\phn$-$6.9 \\
1820.887 &  0.122 &    \phn\phs63.4 &  \phn$-$0.2 &        $-$124.7 & 
\phn$-$6.7 \\
1821.864 &  0.236 &       \phs148.2 &    \phs15.8 &        $-$190.4 & 
\phn$-$1.9 \\
1822.865 &  0.353 &       \phs120.9 & \phn\phs3.7 &        $-$171.9 &
\phn\phs1.1 \\
1823.814 &  0.464 &    \phn\phs49.2 & \phn\phs0.0 &        $-$105.9 & 
\phn$-$2.9 \\
1823.924 &  0.476 &    \phn\phs35.7 &  \phn$-$3.4 &     \phn$-$95.7 & 
\phn$-$2.8 \\
1824.825 &  0.582 &     \phn$-$45.5 & \phn\phs4.8 & \phn\phn\phs1.0 &
\phn\phs2.0 \\
1824.950 &  0.596 &     \phn$-$54.8 & \phn\phs8.4 &    \phn\phs16.6 &
\phn\phs4.5 \\
1830.842 &  0.284 &       \phs142.9 & \phn\phs7.5 &        $-$187.7 &
\phn\phs3.8 \\
1830.946 &  0.296 &       \phs136.5 & \phn\phs2.5 &        $-$180.0
&    \phs10.1 \\
1888.797 &  0.045 &     \phn$-$29.1 &  \phn$-$4.5 &         $-$34.0 & 
\phn$-$6.5 \\
1889.746 &  0.156 &    \phn\phs91.2 &  \phn$-$2.1 &        $-$148.5 & 
\phn$-$0.0 \\
1892.743 &  0.506 &         \nodata &     \nodata &         \nodata
&     \nodata \\
1893.774 &  0.626 &     \phn$-$77.2 &    \phs11.8 &    \phn\phs25.5
&     $-$13.0 \\
1894.771 &  0.742 &        $-$168.5 & \phn\phs4.5 &       \phs131.3 &
\phn\phs6.6 \\
1894.845 &  0.751 &        $-$183.3 &  \phn$-$6.0 &       \phs136.3 &
\phn\phs7.2 \\
1895.721 &  0.853 &        $-$196.3 &  \phn$-$1.8 &       \phs144.7 & 
\phn$-$2.1 \\
1895.811 &  0.864 &        $-$184.3 & \phn\phs7.5 &       \phs144.7 &
\phn\phs0.6 \\
1896.651 &  0.962 &        $-$132.7 &  \phn$-$8.0 &   \phn\phs 76.7 &
\phn\phs1.4 \\
1896.784 &  0.977 &        $-$103.8 & \phn\phs4.1 &   \phn\phs 45.3
&     $-$12.6 \\
1897.652 &  0.078 & \phn\phn\phs6.0 &     $-$10.1 &     \phn$-$71.1 & 
\phn$-$1.8 \\
1897.784 &  0.094 &    \phn\phs21.0 &     $-$12.9 &     \phn$-$87.4 &
\phn\phs0.1 \\
1898.656 &  0.196 &       \phs114.5 &  \phn$-$3.8 &        $-$177.2 & 
\phn$-$3.2 \\
1898.793 &  0.212 &       \phs126.0 & \phn\phs0.8 &        $-$173.9 &
\phn\phs7.2 \\
1899.658 &  0.312 &       \phs133.2 & \phn\phs2.5 &        $-$182.4 &
\phn\phs4.3 \\
1899.791 &  0.328 &       \phs124.9 &  \phn$-$1.5 &        $-$182.2 &
\phn\phs0.2 \\
1900.653 &  0.429 &    \phn\phs69.7 &  \phn$-$4.9 &        $-$136.9 & 
\phn$-$7.7 \\
1900.784 &  0.444 &    \phn\phs60.6 &  \phn$-$3.2 &        $-$124.9 & 
\phn$-$6.7 \\
1901.637 &  0.543 &     \phn$-$32.7 &     $-$15.8 &     \phn$-$14.1 &  
\phs 21.2 \\
1901.770 &  0.559 &     \phn$-$27.6 & \phn\phs3.0 &     \phn$-$24.4 & 
\phn$-$3.0 \\
\enddata
\end{deluxetable}

\clearpage


\begin{deluxetable}{lr}
\tablewidth{0pc}
\tablecaption{Orbital Elements for HD 17505Aa \label{17505orb}}
\tablehead{
\colhead{Element} &
\colhead{Value} }
\startdata
$P$~(days)                  \dotfill & 8.5710 (8) \\
$T$ (HJD-2,450,000)         \dotfill & 1862.696 (16) \\
$e$                         \dotfill & 0.095 (11) \\
$\omega$ ($^\circ$)         \dotfill & 252 (6) \\
$K_1$ (km s$^{-1}$)         \dotfill & 166.5 (18) \\
$K_2$ (km s$^{-1}$)         \dotfill & 170.8 (18) \\
$M_2/M_1$                   \dotfill & 0.975 (15)  \\
$\gamma_1$ (km s$^{-1}$)    \dotfill & $-25.8$ (12) \\
$\gamma_2$ (km s$^{-1}$)    \dotfill & $-26.3$ (12) \\
$M_1$ sin$^{3}i$ ($M_\odot$)\dotfill & 17.1 (6) \\
$M_2$ sin$^{3}i$ ($M_\odot$)\dotfill & 16.6 (6) \\
$a$ sin $i$ ($R_\odot$)     \dotfill & 56.8 (4) \\
r.m.s.$_1$ (km s$^{-1}$)    \dotfill & 7.0 \\
r.m.s.$_2$ (km s$^{-1}$)    \dotfill & 7.0 \\
\enddata
\tablecomments{Numbers in parentheses give the error in the last digit
quoted.}
\end{deluxetable}

\clearpage



\clearpage

\clearpage 

\clearpage

\clearpage

\clearpage

\clearpage

\clearpage

\clearpage

\end{document}